\documentclass[fleqn,10pt]{wlscirep}
\usepackage[utf8]{inputenc}
\usepackage[T1]{fontenc}
\usepackage{bm}

\usepackage[framemethod=TikZ]{mdframed}
\usepackage{xcolor}
\usepackage{addfont}
\addfont{OT1}{capbas}{\capbas}
\addfont{OT1}{capbasd}{\capbasd}

\newmdenv[%
    backgroundcolor=blue!8,
    linecolor=blue,
    outerlinewidth=1pt,
    roundcorner=5mm,
    skipabove=\baselineskip,
    skipbelow=\baselineskip,
]{NewBox}

\newcommand{\ClearlyBox}[1]{
\begin{NewBox}
#1
\end{NewBox}}

\title{
Gravitational-Wave and X-ray Probes of the Neutron Star Equation of State
}

\author[1,*]{Nicol\'as Yunes}
\author[2,**]{M. Coleman Miller}
\author[3,***]{Kent Yagi}

\affil[1]{Illinois Center for Advanced Studies of the Universe, Department of Physics,
University of Illinois Urbana-Champaign, Urbana, IL 61801, United States}
\affil[2]{Department of Astronomy and Joint Space-Science Institute, University of Maryland, College Park, MD 20742-2421, United States}
\affil[3]{Department of Physics, University of Virginia, Charlottesville, Virginia 22904-4714, United States}

\affil[*]{e-mail: nyunes@illinois.edu}
\affil[**]{e-mail: miller@astro.umd.edu}
\affil[***]{e-mail: ky5t@virginia.edu}

\begin{abstract}

Neutron stars are a remarkable marriage of Einstein's theory of general relativity with nuclear physics. Their interiors harbor extreme matter that cannot be probed in the laboratory. At such high densities and pressures, their cores may consist predominantly of exotic matter such as free quarks or hyperons. Gravitational wave observations from the Laser Interferometer Gravitational-wave Observatory (LIGO) and from other interferometers, and X-ray observations from the Neutron Star Interior Composition Explorer (NICER), are beginning to pierce through the veil. These observations provide information about neutron star cores, and therefore, about the physics that makes such objects possible. In this review, we discuss what we have learned about the physics of neutron stars from gravitational wave and X-ray observations. We focus on what has been observed with certainty and what should be observable in the near future, with an eye out for the physics that these new observations will teach us. 

\end{abstract}
\begin{document}

\flushbottom
\maketitle

\thispagestyle{empty}
	
\noindent \textbf{Key points:} 
\begin{itemize}
    \item The processes at play inside neutron stars encode general relativity, quantum mechanics, particle physics and nuclear physics that cannot be replicated in the lab. 
    \item Gravitational wave observations of binary neutron star mergers are beginning to provide information about the equation of state of supranuclear matter through constraints on the tidal deformability of neutron stars. 
    \item The X-rays emitted by hot spots on the surface of certain pulsars are beginning to provide information about nuclear physics through constraints on the radius of neutron stars. 
    \item Future observations of gravitational waves and X-rays from LIGO, Virgo, KAGRA and NICER will provide unprecedented insights into the physics of neutron stars.
\end{itemize}

\noindent \textbf{Website summary:} The observation of gravitational waves emitted in the merger of neutron stars and the observations of X-rays emitted by hot spots on their surfaces are beginning to reveal nuclear physics insights about these compact objects.  

\section{The marvelous neutron star}

Neutron stars are wonderfully marvelous. With masses comparable to our Sun's, but radii of only approximately 12 km, they are some of the most compact, and thus gravitationally powerful objects in the universe~\cite{Shapiro:1983du}. With monstrous magnetic fields \cite{Kaspi:2017fwg}, sometimes a trillion times stronger than that of a refrigerator magnet, they funnel photons into beams that travel astronomical distances. 
With astounding rotation speeds that can reach up to hundreds of Hertz, rivaling professional kitchen blenders, they whip these magnetic fields and beams around, creating astrophysical lighthouses. Every time the beams cross Earth, radio telescopes record a pulse, and the counting of these pulses creates a clock. Neutron stars are, in fact, one of the most stable and accurate clocks known in the natural universe because of their marvelously stable rotation rates~\cite{doi:10.1126/science.238.4828.761}.

Neutron stars are also wonderfully unavoidable. A massive star is supported against gravity during most of its life by the radiation force that it produces in the thermonuclear fusion of its component gases. The by-products of this reaction are ever heavier elements, and thus, at some point in the star's lifetime, it is no longer energetically favorable to continue the burning sequence. The massive star then sheds its outer layers in a supernova explosion, leaving behind an iron core that can no longer burn, and thus can no longer prevent gravitational collapse. In the iron core goes, just as dictated by Einstein's theory of general relativity, and so if nothing came to the rescue, one would expect the formation of a black hole. 

Neutron stars, however, are wonderfully quantum mechanical. As the iron core contracts, it becomes energetically favorable for electrons and protons to combine to form neutrons and emit neutrinos via inverse beta decay. The iron core has now become essentially a soup of neutrons, peppered of course with a few other particles. But neutrons are fermions, and this soup is so dense that the distance between fermions becomes really small, forcing each neutron to feel the influence of its neighboring neutrons. By the Pauli exclusion principle, this then leads to a very large (Fermi) momentum and energy, and thus, a very large ``degeneracy'' pressure, whose gradient halts the collapse. 

\begin{figure}[ht]
\centering
\includegraphics[width=9.5cm]{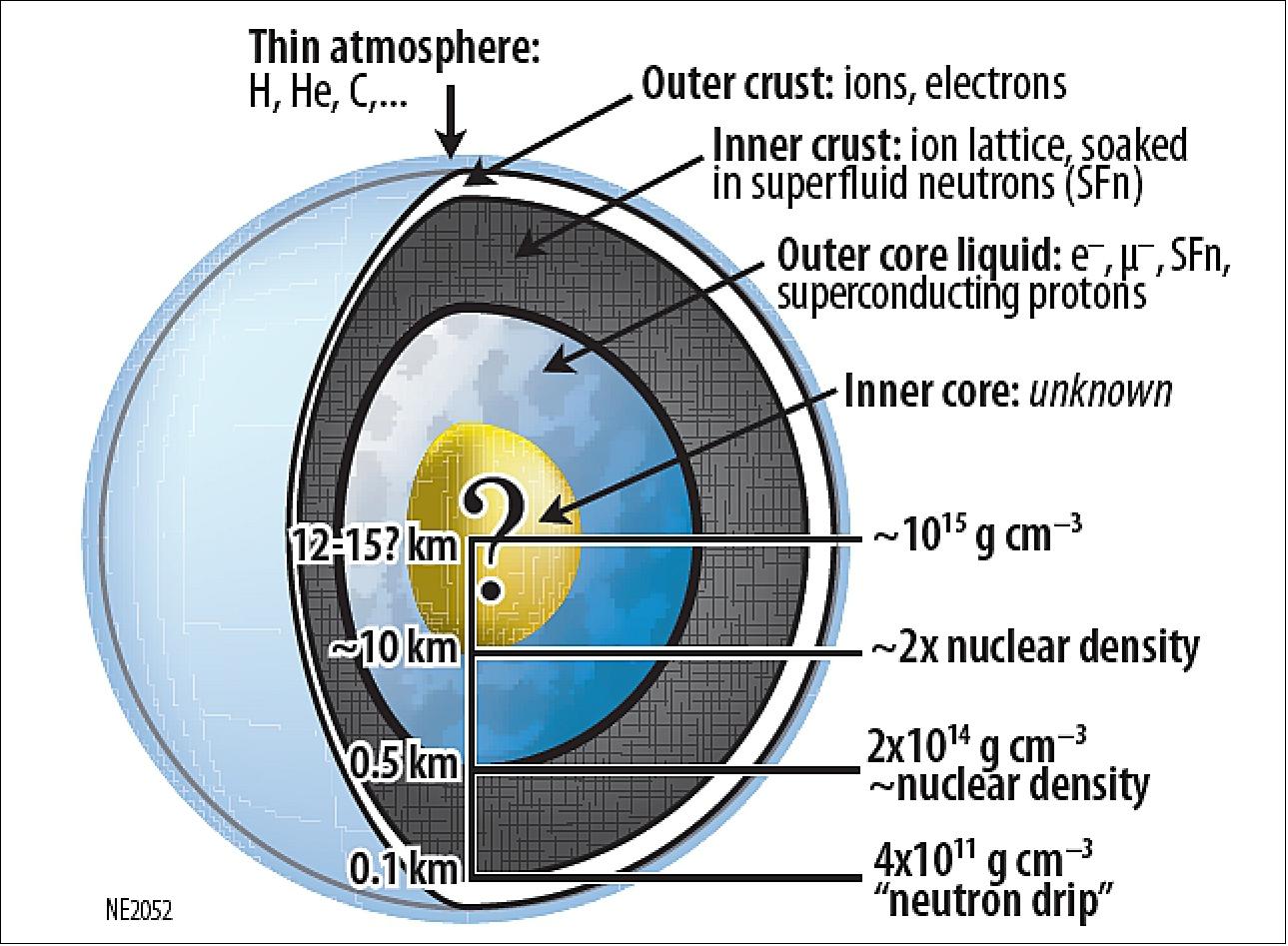}
\caption{Cartoon structure of a neutron star and its internal structure.
}
\label{fig:EoSStruc}
\end{figure}

As a result, neutron stars are wonderfully dense.  Although their atmospheres (which comprise only the few centimeters closest to the surface) can have atoms (albeit possibly distorted into near-cylinders by the strong magnetic field), just a bit deeper, matter is so dense that electrons do not belong to individual nuclei.  At densities greater than $\sim 10^7$~g~cm$^{-3}$, the Fermi energy of electrons becomes high enough that the matter becomes progressively richer in neutrons~\cite{Baym:1971pw}, which produces nuclei such as $^{120}$Rb, which have 40 protons and 80 neutrons.  At densities greater than $4\times 10^{11}$~g~cm$^{-3}$ it becomes possible for neutrons to ``drip out'' of the nucleus,
which means that matter is a mix of free neutrons, free electrons, and nuclei.   At even higher densities,  nuclei can cluster together to form ``pasta-like'' structures, such as one-dimensional strings (``spaghetti") or two-dimensional surfaces (``lasagna"). Going in a bit deeper, at about ``nuclear saturation density'' ($2.7\times 10^{14}$~g~cm$^{-3}$, so called because it corresponds to the density at the center of large nuclei), there are no longer any isolated nuclei, and we now have the neutron soup mentioned above, as shown in Fig.~\ref{fig:EoSStruc}. Pushing to yet higher densities, it may be energetically favorable to form other baryons with at least one strange quark, such as hyperons~ \cite{Ambartsumyan,Chatterjee:2015pua}, until eventually close to the center of the star, quarks may become deconfined~\cite{Ivanenko:1965dg} and one may encounter a degenerate quark-gluon plasma.

Neutron stars are therefore wonderfully cool, but not just because they are physically interesting: they are also ``cold.'' Imagine a neutron star composed of an immense number of fermions, which due to the Pauli exclusion principle must occupy different energy states. The highest energy state, the Fermi energy, is therefore enormous, and one can think of these fermions as forming a kind of ``gas.'' If so, you can associate an effective temperature to this gas, by dividing the Fermi energy by the Boltzmann constant. This temperature turns out to be remarkably high, on the order of $10^{12}$~K, about 5 orders of magnitude higher than the temperature at the center of the Sun. Therefore, even though isolated neutron stars are actually very hot in terms of their actual temperature, which at their cores might typically be $10^8-10^{10}$~K, they are cold relative to their Fermi temperature. Of course, when neutron stars are born, they can be significantly hotter; for example, proto-neutron stars can have core temperatures of $10^{11}$~K. But such stars cool down very rapidly via neutrino emission, dropping by two orders of magnitude in just a thousand years, which is an extremely short time scale by astronomical standards. Thus, unless one is observing their birth, neutron stars are, for most purposes, cold objects relative to their Fermi temperature.

Although it may not seem so from the above description of their complicated structure, there are some ways in which neutron stars are wonderfully simple.  Many macroscopic aspects of neutron stars that have been or could be observed using electromagnetic radiation or gravitational waves, including their mass, radius, and tidal deformability, depend only on the nature of gravity (here we assume general relativity) and the \emph{equation of state}~\cite{lattimer_prakash2001}, which determines the pressure given other quantities such as the energy density, temperature, and composition.  As argued above, the temperature is too low to make a difference in isolated and ``old'' neutron stars, although temperature cannot be neglected in the merger of neutron stars.  The composition is usually assumed to be the equilibrium composition, because there is enough free energy to transition to the ground state (unlike, say, at terrestrial densities, where $^{56}$Fe is the ground state of matter but there is insufficient energy to cause fusion to guarantee that matter reaches that state).  Other complications are also thought to be ignorable in the description of the equation of state; for example, shear and bulk viscosity are believed to contribute negligibly to the equation of state (although they may be relevant in the dynamics of the merger of neutron stars~\cite{Most:2021zvc}). Thus, to an excellent approximation, in neutron star cores the pressure depends only on the energy density, which is to say that the equation of state is \emph{barotropic}.

\ClearlyBox{
\begin{center}
    {\bf{What is the equation of state, and how can it be inferred from observations of neutron stars?}} 
\end{center}
\vspace{0.2cm} 
As indicated in the text, the equation of state represents a mapping between the pressure and other quantities, such as the energy density, temperature, and composition. But in neutron stars, it is typically thought that in the core the pressure $p$ depends only on the energy density $\epsilon$, so $p=p(\epsilon)$.  To see how observations of neutron stars can help constrain the equation of state, note that the equation of hydrostatic equilibrium, for nonrotating (and thus spherical) fluid stars in general relativity, is the Tolman-Oppenheimer-Volkoff (TOV) equation~\cite{Shapiro:1983du}
\begin{equation}
\frac{dp}{dr}=-\frac{Gm}{r^2}(\epsilon/c^2)\left(1+\frac{p}{\epsilon}\right)\left(1+\frac{4\pi r^3p}{mc^2}\right)\left(1-\frac{2Gm}{rc^2}\right)^{-1}\; .
\end{equation}
Here $r$ is the coordinate distance from the center of the star and $m$ is the gravitational mass inside this radius, while $G$ is Newton's gravitational constant and $c$ is the speed of light. The quantity $\epsilon$ includes the rest-mass density, so in the Newtonian limit where $p\ll\epsilon$, $\epsilon$ is dominated by rest mass-energy $\rho c^2$. In this same limit, $2Gm/c^2\ll r$, so the above equation reduces to the Newtonian hydrostatic equilibrium equation $dp/dr=-\rho(Gm/r^2)$.  Given an equation of state $p(\epsilon)$ and a central density $\epsilon_c$, one can integrate this equation, combined with the continuity equation $dm/dr=4\pi r^2\epsilon/c^2$, to find the (total) mass $M$ and radius $R$ of a star. In the Newtonian limit, larger $\epsilon_c$ guarantees larger $M$, but in general relativity there is a central density $\epsilon_{c,{\rm max}}$ beyond which further increase leads to lower $M$. This corresponds to an instability, and means that in general relativity a given equation of state has a maximum stable mass.  Thus, any measurement or constraint on the mass and radius of a star (or other quantities such as the mass and tidal deformability of a star, or the mass and moment of inertia of a star) can be compared with the predictions of a set of equations of state.

In these neutron star structure equations, however, there is no information about the composition of the core.  Any composition that yields a $p(\epsilon)$ that is consistent with observations will do.  Thus, although observations of neutron star mass, radius, tidal deformability, moment of inertia, etc. provide valuable constraints, they cannot by themselves tell us whether the core is mainly neutrons, or hyperons, or free quarks, or something else.  Some additional information can in principle be obtained by measurements of the temperatures of neutron stars of a given age and mass (because temperatures depend on transport properties that in turn have some dependence on composition) or about the gravitational waves emitted in the (hot) merger of neutron star binaries. However, at this time, temperature and merger information is a bit too uncertain or unavailable to provide strong and reliable constraints.
}

\section{The equation of state puzzle}

The problem is then ``simple'': solve for the equation of state using many-body quantum mechanics and quantum chromodynamics, and then use this in conjunction with the Einstein equations to predict the observable properties of the equation of state. Unfortunately, this is easier said than done. The Einstein equations part is not the problem. In fact, it is relatively straightforward to compute the observable properties of neutron stars in general relativity, once you are given an equation of state. The problem is quantum mechanical, and relates to the ``sign problem"~\cite{Troyer:2004ge} present in quantum chromodynamics calculations at nonzero baryon chemical potential. Although this theory has straightforwardly calculable predictions\cite{Aoki:2006we} for matter at finite temperature and zero net baryon density (i.e., matter with almost equal numbers of baryons and antibaryons), current numerical methods become exponentially more costly when the net baryon density is large. As a result, it is not possible to compute the properties of neutron star core matter using current first principles approaches.

For this reason, an abundance of models for the equation of state of matter at supranuclear densities and effectively zero temperature have been put forth~\cite{Baym:2017whm}. 
One approach is to solve quantum-chromodynamic-motivated models using the best microphysics possible given the constraints of the sign problem, see e.g.~\cite{Nambu:1961tp,Alford:1997zt,Dexheimer:2009hi,Tews:2012fj}.
Another is to assume that we know the equation of state up to some threshold density (which is typically around 
nuclear saturation density) and then extrapolate to higher densities using some (often parameterized) one-dimensional function, such as a piecewise polynomial~\cite{Read:2008iy}.  In this latter approach, the aim is to use nuclear physics experiments and astrophysical observations to constrain this phenomenological function, and then to study what the constraints imply for nuclear physics~\cite{Annala:2017llu}.    

\begin{figure}[ht]
\centering
\includegraphics[width=15cm]{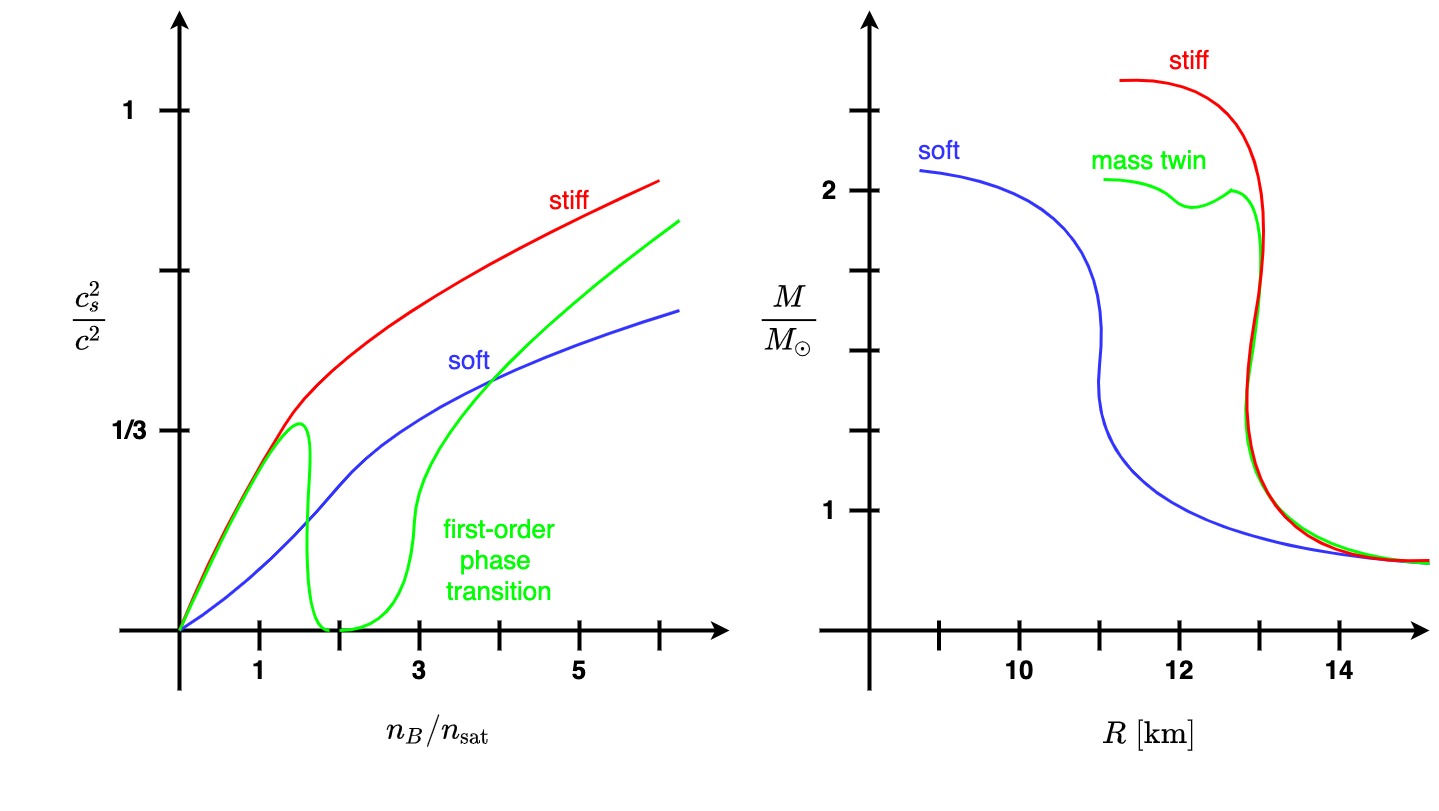}
\caption{Schematic representation of equations of state showing the square of the sound speed against baryon number density normalized to nuclear saturation density (left) and their corresponding mass-radius curves for neutron stars (right). Generally, stiff equations of state have larger sound speed, maximum mass and radius than soft ones. Equations of state with first-order phase transition (energy densities at which $c_s^2 = 0$) can produce mass twins if they are large enough and if they occur at densities around what is shown in the figure.
}
\label{fig:EoS-M-R}
\end{figure}

The phenomenological nature of the second approach does not imply that interesting nuclear and particle physics is unimportant. On the contrary, the interactions between nuclei, neutrons and quarks can greatly influence the functional form of the equation of state, and therefore, both approaches attempt to include as much physics as possible. For example, if the matter in the core of neutron stars transitions into deconfined quarks, then under certain circumstances it is possible for the speed of sound of the fluid (the square root of the derivative of the pressure with respect to the energy density) to become very small or zero~\cite{Alford:2013aca}. This is called a first-order phase transition in the quantum chromodynamic phase diagram, which for old and isolated neutron stars is two dimensional (pressure versus density or chemical potential only). The speed of sound also cannot exceed the speed of light (the so-called ``causal limit''), and at extremely high energy densities it is expected to approach square root of one third the speed of light (the so-called ``conformal limit'')~\cite{Haque:2014rua}. The latter arises because at sufficiently high densities, the particles' energy is dominated by their Fermi momentum, so they can be treated effectively as a relativistic gas. It is not currently known if the conformal limit applies at the densities expected inside the cores of neutron stars, or whether other phase transitions may be present. 
The equation of state puzzle then may be solved through observations, since certain quantities that are (directly or indirectly) observable are determined by the equation of state (see the box for more details). 
Equations of state are sometimes called ``stiff'' or ``hard" when the slope of the pressure-energy density curve is large. On the other hand, a ``soft" equation of state, for which the pressure increases slowly as the energy density increases, has a smaller maximum mass. Equations of state that contain first-order phase transitions around nuclear saturation density allow for stars with similar masses but different radii, which have been dubbed \emph{mass twins}~\cite{Glendenning:1998ag}. Figure~\ref{fig:EoS-M-R} shows a schematic representation of stiff and soft equations of state, and equations of state with first order phase transitions, together with their respective mass-radius curves.  Other interesting properties of nuclear/quark matter include heat capacity, superfluidity, viscosity, and shear modulus that may be probed through cooling~\cite{Page:2004fy,Blaschke:2004vq}, glitches~\cite{Piekarewicz:2014lba,Haskell:2015jra}, and r-mode oscillations~\cite{2018arXiv180404952F,2021arXiv210403137H} of neutron stars. Other basic physics questions that one could address through neutron star observations include   the role of three-body forces and the threshold for the appearance of hyperons (the ``hyperon puzzle'').  
    
\section{A new dawn for nuclear astrophysics}

Up until the 2010s, most astrophysical information about neutron stars came from observations of the light they emit. Perhaps the best known of these are the radio pulses emitted by neutron stars in a binary orbit with another compact companion, such as another neutron star or a white dwarf. The timing of these pulses allow the careful reconstruction of the binary orbits, including relativistic effects, which earned Hulse and Taylor~\cite{Hulse:1974eb} a Nobel Prize in Physics in 1993. Among the many wonders these binary pulsars have revealed, what concerns us the most here is the inference of the masses of the binary companion. To date, the heaviest neutron star with a well-measured mass is PSR~J0740+6620 (where the numbers represent the location of the source in the sky in astronomical notation) at $m=2.08\pm 0.07~M_{\odot}$~\cite{Cromartie:2019kug}. Since high neutron star masses can be compared with the maximum mass predicted using different equations of state, the observation of  PSR~J0740~\cite{Cromartie:2019kug} and similarly high-mass neutron stars~\cite{1.97NS,2.01NS} rules out equations of state that are too soft. Indeed, neutron star mass measurements based on radio timing have been, and continue to be, a lynchpin for astronomical constraints on matter beyond nuclear density, because the data, analysis, and model inferences are all well-understood.
 
A new window to the universe and to neutron stars opened in 2015 when the advanced Laser Interferometer Gravitational-wave Observatory (or advanced LIGO for short) made the first direct detection of gravitational waves~\cite{Abbott:2016blz}. These waves are perturbations to gravity predicted by Einstein and produced when massive objects accelerate. Because of the feebleness of the gravitational force, a tremendous amount of mass has to accelerate to a tremendous degree for us to be able to observe gravitational waves that originate a cosmological distance away from Earth. But this is exactly what happened on 14 September 2015. The GW150914 event (named after its discovery date) was shown to be produced by the merger of two black holes with masses roughly 30 times that of our Sun, at velocities close to half the speed of light~\cite{Abbott:2016blz}. This single event announced the birth of gravitational wave astrophysics. 

Many discoveries have been made in gravitational waves since 2015~\cite{LIGOScientific:2020ibl}, but the most exciting for us was made in 2017~\cite{TheLIGOScientific:2017qsa}. On August 17th of that year, advanced LIGO (and this time, also another gravitational wave detector called Virgo, too) detected gravitational waves again, but this time the frequency at which the signal peaked was much higher than in 2015.  For objects such as black holes and neutron stars, whose radius $R$ is just a few times $Gm/c^2$, a higher frequency is a tell-tale sign of a much lower-mass merger.  This is because Kepler's Third law tells us that the square of the orbital frequency, which is directly proportional to the square of the gravitational wave frequency, scales linearly with the total mass of the binary and inversely with the separation $d$ cubed. Thus at merger, when $d\sim R\sim G m/c^2$, the gravitational wave frequency is inversely proportional to the binary's total mass. A detailed analysis of the GW170817 event later revealed that it was likely produced by the coalescence of two neutron stars, with masses of $\sim 1.3-1.4~M_\odot$  at a mere 40 Mpc (i.e., 130 million light years) away from Earth~\cite{TheLIGOScientific:2017qsa}. On a human scale, that is a ridiculous distance (astronomical in fact!), but to an astrophysicist, this is incredibly close (it corresponds to a cosmological redshift of about $0.009$.) The closer the event, the stronger the signal, so using this event it was possible for the first time to extract information about the equation of state from gravitational waves.  

\begin{figure}[ht]
\centering
\includegraphics[width=9.5cm]{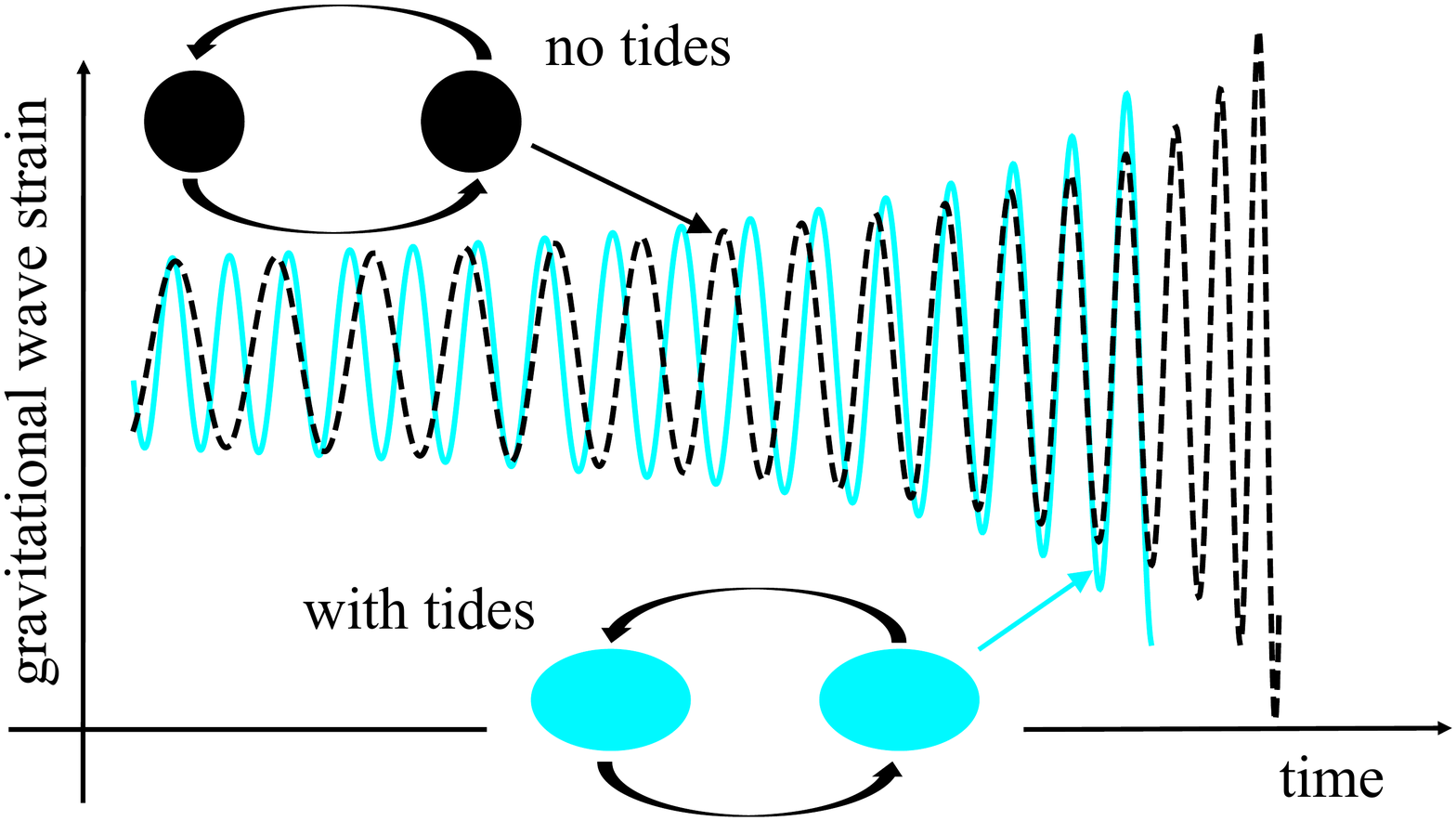}
\caption{ Gravitational waveform from the last few cycles of a compact binary inspiral with (cyan solid) and without (black dashed) the effect of tides. The former (latter) corresponds to waves generated by inspiraling black hole (neutron star) binaries with vanishing (non-vanishing) tidal deformabilities. The waveform data is taken from~\cite{Carson:2020gwh}.}
\label{fig:coalesce}
\end{figure}

But how do gravitational waves carry information about the equation of state? When two neutron stars spiral into each other and collide, before the collision takes place they are tidally perturbed by each other's gravitational field. Just like Earth acquires a tidal bulge due to the Moon, inducing high and low tides in the ocean, when a neutron star gets close to another compact object, it will tidally deform. This tidal deformation requires energy, and so the neutron star ``borrows'' it from the orbital energy, therefore forcing the binary to spiral in faster than it would have otherwise. A speed up in the rate of inspiral directly affects how the gravitational wave frequency changes with time, because, as we said before, the orbital and gravitational wave frequencies are linearly related. Therefore, by carefully monitoring the evolution of the gravitational wave phase, one can in principle extract information about how much the objects that produced the wave were tidally deformed on their way to coalescence~\cite{flanagan-hinderer-love,hinderer-lackey-lang-read} (see Fig.~\ref{fig:coalesce}).

This is exactly what the LIGO Scientific and Virgo collaborations measured from the GW170817 event~\cite{TheLIGOScientific:2017qsa,LIGOScientific:2018cki,Chatziioannou:2020pqz}. The signal to noise ratio was large enough that, from the gravitational wave data alone, a double neutron star merger in which both stars had the same equation of state was somewhat preferred over other options, including two black holes (which is also strongly disfavored by the subsequent electromagnetic emission) and one neutron star and one black hole.  If the event involved a binary neutron star system similar in its spins to those we observe in the Milky Way, then the probability distribution of the tidal deformability for both stars peaks at a nonzero value and implies a radius between $\sim 10.5$~km and $\sim 13.5$~km for both stars, at 90\% credibility~\cite{LIGOScientific:2018cki}. 

\section{From gravitational waves to dense matter}

But how on Earth do you constrain the radius of neutron stars from a measurement of the tidal effect on the gravitational waves emitted? In very broad terms, there are essentially two approaches that have been pursued to do so.  

The standard approach is to adopt some model for the equation of state~\cite{Read:2008iy,Lindblom:2010bb,Greif:2018njt,Tews:2018kmu} and then carry out Bayesian parameter estimation. In general, all equation of state models contain some low-density part, which is then extrapolated in some specific mathematical way to higher densities beyond some threshold density (which has been typically taken to be between half and twice nuclear saturation density). One then employs standard Bayesian inference: for a given draw of the equation of state model and two draws of the central densities, one predicts the masses of the binary companions and their tidal deformabilities, and from this, the gravitational wave model that one compares to the gravitational wave data.  Based on the relative likelihoods of the various draws in the exploration of the likelihood surface, one can then obtain posterior distributions for the pressure versus energy density, the mass versus radius, or other quantities. 

The main caveat of this approach stems from the need to prescribe an equation of state model. Any given model will place greater prior weight on some portions of equation of state parameter space than on others. For example, some models may exclude the possibility of first-order phase transitions,
whereas others might strongly emphasize them. Some models may exclude wiggles, kinks and other crossover-type structure in the speed of sound, while others might focus on them. A reasonable approach then is to use a few distinct models in the hope (which fortunately is more and more the reality) that the data are informative enough that distinct but reasonable models lead to similar posteriors on the mass, radius and other observables.

Another complementary approach relies on relations (often called ``universal relations") between neutron star properties that are insensitive to the equation of state~\cite{Yagi:2016bkt}.  Particularly tight relations of this type include the so-called I-Love-Q relations~\cite{Yagi:2013bca,Yagi:2013awa} between the moment of inertia (I), the tidal deformability or Love number (Love) and the quadrupole moment (Q).  For the purpose of inferring the radius of neutron stars from gravitational wave data, the most important universal relation is the \emph{binary Love} one~\cite{Yagi:2015pkc,Yagi:2016qmr} (see~\cite{De:2018uhw} for a similar relation), which relates the anti-symmetric combination of the tidal deformabilities to their symmetric combination and the mass ratio of the binary, and the \emph{Love-C} one~\cite{Maselli:2013mva,Yagi:2016bkt}, which relates the tidal deformability of a star to its compactness. The binary Love relations can be used to effectively reduce the number of independent parameters needed to describe the gravitational waveform, and thus, improve the precision with which the tidal deformabilities can be measured~\cite{Yagi:2015pkc,Yagi:2016qmr,Chatziioannou:2018vzf}. With a posterior on the deformabilities, the Love-C relation then provides the compactness of each star, which when combined with the gravitational wave measurement of the binary component masses, yields the radius of each star. 

\ClearlyBox{
\begin{center}
    {\bf{How do you use relations insensitive to the equation of state to learn about the latter? \\ And what is the Love number anyways?}} 
\end{center}
\vspace{0.2cm} 
Let us begin with the second question first. The Love numbers are a set of real numbers introduced by Augustus Edward Hough Love in the 1900s to describe the Earth tides caused by the Moon. But of course one can study the tides of any massive body caused by any external perturbation, including the tides of a neutron star due to its binary companion. The Love numbers are proportional to the tidal deformability, which describes how much a star deforms in response to an external perturbation~\cite{hinderer-love,binnington-poisson,damour-nagar}. More precisely, an external perturbation generically induces a redistribution of the isodensity contours inside a massive body, which in turn can be described through mass and current multipole moments of the mass distribution. The tidal deformabilities are then formally defined as the constants of proportionality that relate how much of a multipolar deformations is induced in a star ${\cal{M}}_{\ell}$ due to an external tidal perturbation ${\cal{P}}_{\ell}$, i.e.~${\cal{M}}_{\ell} = \lambda_{\ell} {\cal{P}}_{\ell}$.

The external perturbation can be of two classes (even or ``electric'' and odd or ``magnetic'' parity), depending on how it transforms under parity. Each class, in turn, can be decomposed into multipole moments, with the leading-order perturbation produced by the electric-type quadrupole tidal tensor ${\cal{E}}_{ij}$. Given this, the tidal deformability can also be classified and decomposed in an analogous manner, with the leading-order tidal deformability being the electric-type quadrupole tidal deformability $\lambda_{E,\ell=2}$ or just $\lambda$ for short. This deformability is then defined via the relation $Q_{ij} = \lambda {\cal{E}}_{ij}$, where $Q_{ij}$ is the induced mass quadrupole moment. The calculation of any tidal deformability requires the solution to the perturbed Einstein equations for a star that is being deformed by some ``external universe,'' such as a binary companion sufficiently far away. 

The tidal deformabilities of the neutron stars in a binary affect the gravitational waves they emit during their inspiral because they modify the orbital energy and the rate of gravitational wave emission. For an equal-mass binary, the tidal effect on the orbital energy changes the phase by 5.5 times as much as the tidal modification of the gravitational wave emission rate to the leading post-Newtonian order, irrespective of the equation of state. 
As described in the text, the tidal deformations modify the Hamiltonian of the binary system by adding a term of the form~\cite{flanagan-hinderer-love,Racine:2004hs} $\delta H \propto U_{12} (\lambda_1/m_1^5) (m_1/m)^3 U_{12}^5  + 1 \to 2$, where $U_{12} = G m/r_{12}$ is the binary's Newtonian potential, with $m$ the total mass and $r_{12}$ the orbital separation, and $\lambda_1$ is the electric-type, $\ell=2$ tidal deformability of star 1, which scales with its radius to the fifth power. Since a binary system composed of tidally deformed stars has a larger (i.e., less negative) binary binding energy, it takes less time for gravitational waves to drain this energy away, and for the binary to inspiral. Such a modification in the inspiral orbital dynamics imprints directly on the gravitational waves emitted.    

Gravitational wave detectors are more sensitive to the phase of the wave than its amplitude when one carries out parameter estimation by matched-filtering the data with a template model. Because the covariance matrix of the noise is diagonal in the Fourier domain, one typically carries out matched-filtered parameter estimation in frequency space. The Fourier transform of the waveform for inspiraling neutron stars contains the term~\cite{flanagan-hinderer-love,Yagi:2016qmr} $\delta \Psi(f) \propto \left[f(\eta) \bar{\lambda}_s + g(\eta) \bar{\lambda}_a \right] (\pi m f)^{10/3}$, where $f(\eta)$ and $g(\eta)$ are functions of the symmetric mass ratio $\eta = m_1 m_2/m^2$, while 
$\bar{\lambda}_{s,a} = (\Lambda_1\pm \Lambda_2)/2$ are the non-dimensional, symmetric and asymmetric combinations of the tidal deformabilities with $\Lambda_A \equiv \lambda_A/m_A^5$ representing the dimensionless tidal deformability for bodies $A = (1,2)$. Extracting both combinations from the data from this single term in the Fourier phase seems impossible due to degeneracies among them, but there are two ways out~\cite{LIGOScientific:2018cki}.

One option is to choose an equation-of-state model to compute $\bar{\lambda}_{s,a}$ as a function of the mass of the stars and determine the posteriors for the parameters of the model equation of state and the central densities by comparing to the data. Another option is to use equation-of-state insensitive relations. In the latter approach, one uses the binary Love relations~\cite{Yagi:2015pkc,Yagi:2016qmr} to prescribe $\bar{\lambda}_a$ in terms of $\bar{\lambda}_s$ analytically, thus making $\delta \Psi(f)$ a function of only one $\bar\lambda_s$ and $\eta$. One can then carry out parameter estimation to extract both $\bar\lambda_s$ and $\eta$ (because the symmetric mass ratio also appears in other terms of the Fourier phase independent of the tidal deformabilities). Once $\bar\lambda_s$ has been extracted, one can then use the binary Love relations again to extract $\bar\lambda_a$, and from knowledge of both of these combinations one can trivially extract both $\lambda_1$ and $\lambda_2$, without ever choosing an equation-of-state model.    
}

A caveat of this approach is that for the binary Love relation to apply, the two neutron stars in a binary must both be on the primary stable branch; for example, the relation is inapplicable for twin stars if one star is on one stable branch and the other is on a different, higher-density stable branch~\cite{Carson:2019rjx}.  Care also needs to be applied when using the binary Love relation to place \emph{lower} bounds on the tidal deformability, because these lower bounds can be below what is realistic for neutron stars and therefore may be outside the region in which the binary Love relations are valid \cite{Kastaun:2019bxo}.

\begin{figure}[ht]
\centering
\includegraphics[width=8cm,clip=true]{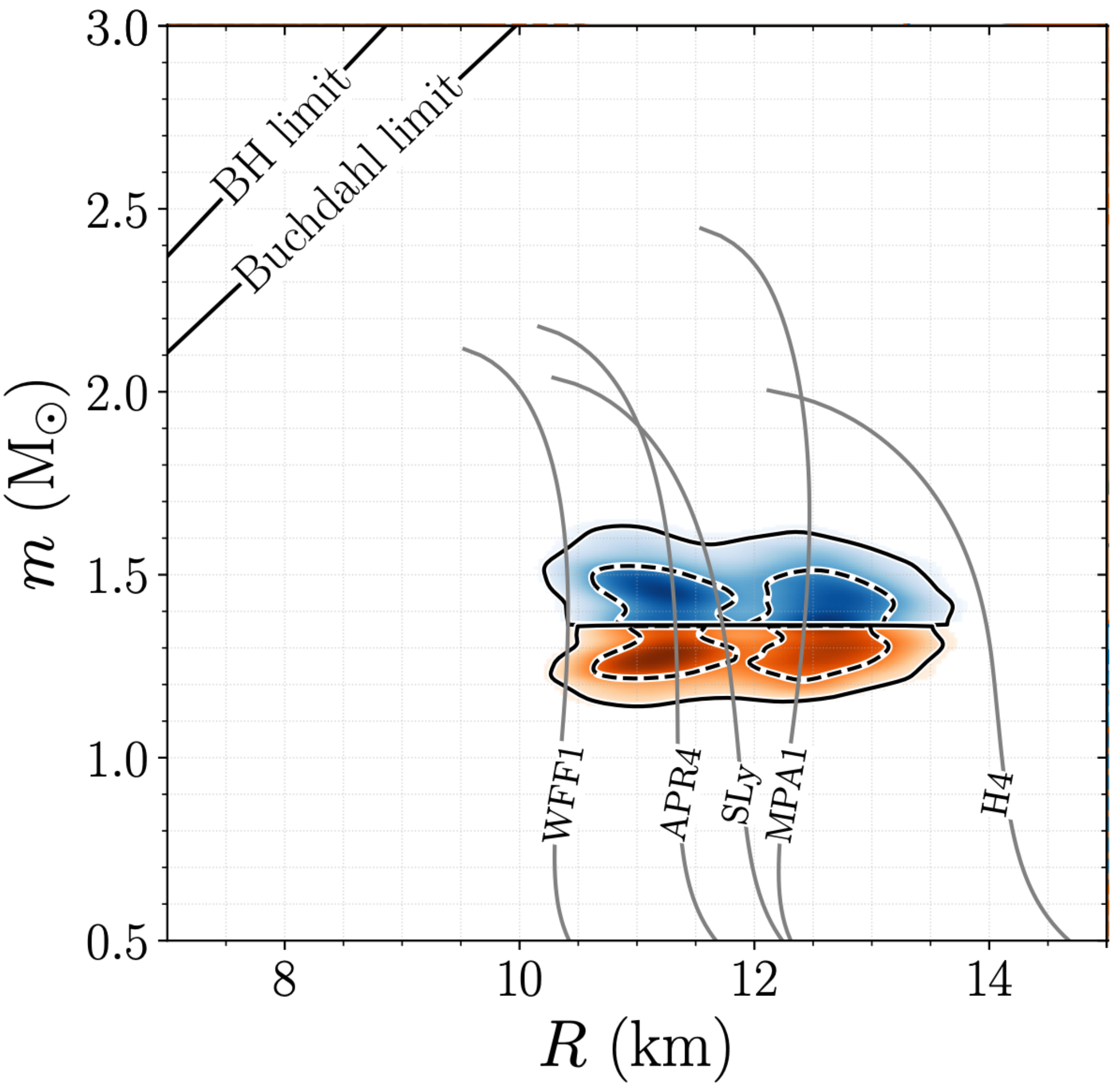}
\includegraphics[width=8cm,clip=true]{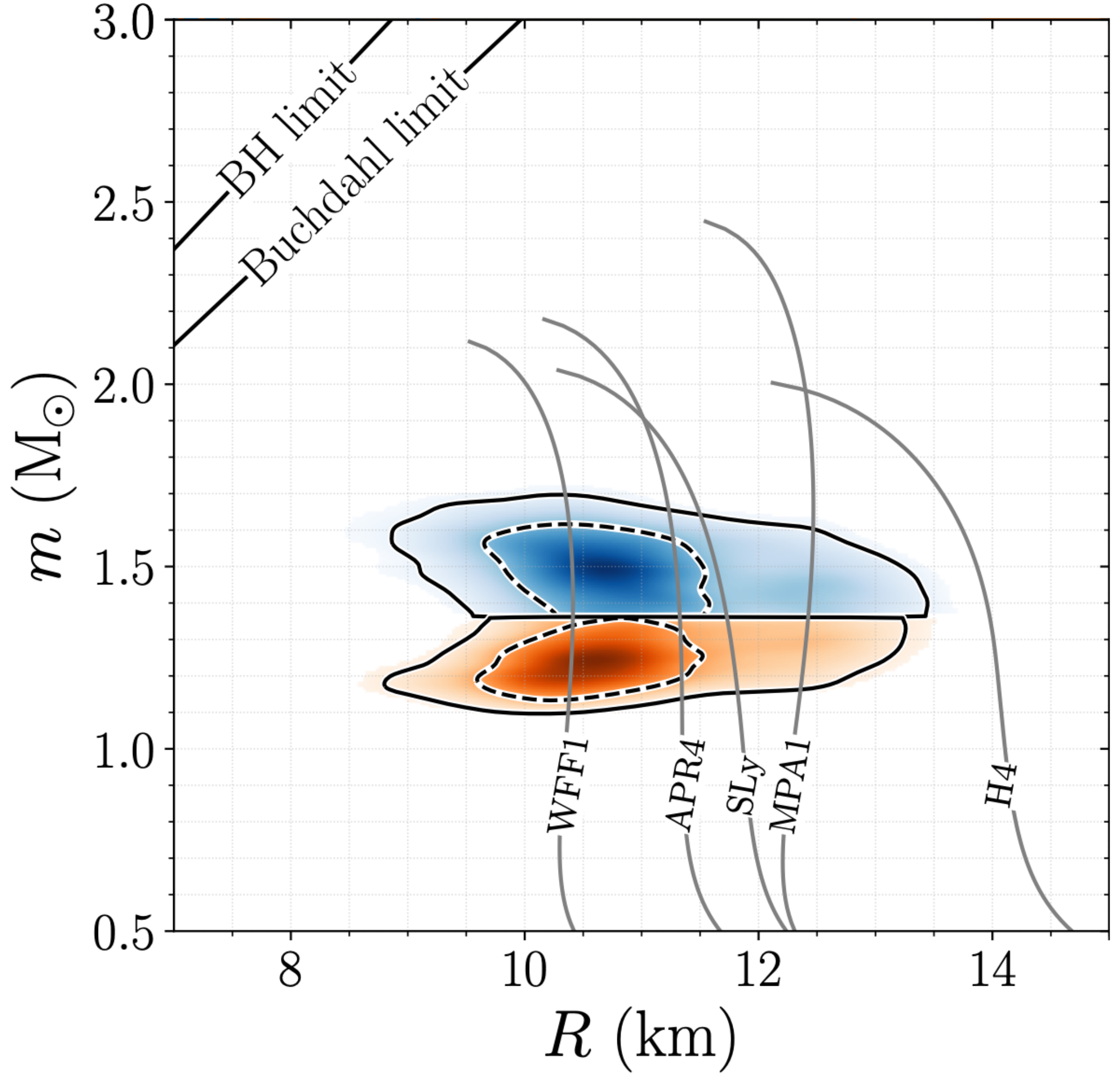}
\caption{ Posterior for the mass and radius of each (blue and orange) components of the binary system that generated GW170817, using a parametrized equation-of-state model (left panel) and the universal-relations method (right panel). The solid and dashed boundaries in each panel correspond to the 90\% and the 50\% credible region. A few mass-radius curves are overplotted in gray for comparison. The solid lines of the top left represent a non-rotating black hole, whose radius is twice its mass (times $G/c^2$ for unit consistency), and the so-called Buchdahl limit~\cite{PhysRev.116.1027} (ie.~the maximum gravitational compactness that static and spherically-symmetric matter configurations must satisfy in general relativity under certain conditions on the energy and pressure). This figure is taken from Ref.~\cite{LIGOScientific:2018cki}. 
}
\label{fig:M-Rbound}
\end{figure}

Both approaches have been applied by the LIGO/Virgo Collaborations on GW170817~\cite{LIGOScientific:2018cki} and the posteriors of the mass and radius are shown in Fig.~\ref{fig:M-Rbound} (see also Ref.~\cite{De:2018uhw} for related work). Not surprisingly, both approaches lead to consistent posteriors in the mass-radius plane in the sense that the 50\% credible regions overlap between the two analyses, suggesting that indeed the data is more informative than any systematic error incurred in the modeling. 

\section{The power of coincidence}

But there is more! The double neutron star coalescence GW170817, which we recall occurred just 40~Mpc away, had counterparts over the entire electromagnetic spectrum~\cite{Monitor:2017mdv}.  Gamma rays indicative of a gamma-ray burst observed $\sim 20-30^\circ$ off-axis were seen just $\sim 1.7$ seconds after the peak of the gravitational wave event.   This was followed by ultraviolet through infrared emission over hours to weeks, with X-rays first detected nine days after the initial event and radio waves visible even now.  

The overall picture is consistent with predictions made prior to the event, in which the energy released by the coalescence of the neutron stars emerges in multiple forms: (1)~gravitational waves, (2)~a short gamma-ray burst (in which the gamma rays are produced by a blast wave moving with a Lorentz factor of at least hundreds), and (3)~far more prolonged emission produced by the quasi-spherical, and much slower ($v\sim 0.01-0.1c$) outflow of unbound matter that was in the neutron stars (where the radiated energy comes from the decay of heavy radioactive nuclei).  This last component, which might have had $\sim 0.01-0.1~M_\odot$ in total mass, has been dubbed a ``kilonova" or ``macronova"~\cite{Metzger:2019zeh}.  The highly neutron-rich outflow is thought to produce heavy elements, such as lanthanides and actinides, with high efficiency; thus, double neutron star mergers, and possibly mergers between black holes and neutron stars, could produce most of the heavy elements in the universe.

GW170817 and events like it provide us with ways to place an upper limit on the maximum mass $M_{\rm max}$ of a non-rotating neutron star, albeit with astrophysical caveats~\cite{2013PhRvL.111m1101B,2015ApJ...812...24F,2015ApJ...808..186L,2017ApJ...850L..19M}. The argument is that if after the merger of two neutron stars the remnant is a long-lived and rapidly rotating neutron star, then if the process of merger generated a strong magnetic field the star would slow down rapidly by magnetic braking.  This would then inject a large fraction of the $\sim 10^{52}$~erg rotational energy into the afterglow and kilonova.  This excess energy is not observed in GW170817, which suggests that instead the merger remnant rapidly collapsed to a black hole.  As a result, the total mass of the binary had to be larger than what can be supported by a rigidly rotating neutron star, and thus the maximum mass is bounded from above.  The caveats are that there is no \textit{direct} evidence for the formation of a black hole (as there would be if the gravitational wave data were many times more precise than they were), and the production of magnetic fields to slow the star's rotation is highly uncertain.  

Nonetheless, under these assumptions, the estimated total mass of the double neutron star binary and the assumption that the remnant collapsed quickly imply that $M_{\rm max}$ for a nonrotating neutron star is less than $\sim 2.2~M_\odot$; note that this relies on a fairly well-understood translation between the maximum mass of a rotating neutron star, such as is formed in the merger, and the maximum mass of a nonrotating star.  Detailed numerical models of the outflow as inferred from electromagnetic observations yield similar answers, with implied values for $M_{\rm max}$ in the range $\sim 2.2-2.3~M_\odot$ \cite{2017PhRvD..96l3012S,2018ApJ...852L..25R,2018PhRvD..97b1501R}.

If these upper limits are reliable, then, in concert with the existence of a few $\sim 2~M_\odot$ neutron stars, they provide a remarkably tight constraint on $M_{\rm max}$: just $\sim 2-2.3~M_\odot$ to be conservative.  This would eliminate both soft equations of state (which have $M_{\rm max}<2~M_\odot$) and hard equations of state (which have $M_{\rm max}>2.3~M_\odot$), therefore cutting down considerably on viable descriptions of the dense matter inside neutron stars.  The most important addition to this information is independent, reliable and precise measurements of neutron star radii, which we now discuss.

\section{A NICER way to nuclear astrophysics}

Precise neutron star radii have long been coveted by nuclear physicists, because they would arguably discriminate between different equation of state models better than any other single measurement.  As a result, numerous radii have been reported over the years, usually focused on X-ray observations of spectra integrated over many neutron star rotation periods.  It has, however, become evident that this method is susceptible to potentially serious systematic errors.  For example, the X-ray emission from a typical cooling neutron star without pulsations can be fit equally well using a pure hydrogen atmosphere, a pure helium atmosphere, and even a black  body spectrum (although neutron stars do not emit as black bodies).  Despite the equally good fits, the inferred radii differ dramatically depending on the assumptions; for example, hydrogen and helium atmospheres can give radii that differ by as much as 50\%.  Thus, even if the formal statistical precision of the radius measurement is excellent, the reliability may not be.

The Neutron star Interior Composition Explorer (NICER) adds a new dimension to the X-ray observations.  In additional to its other observing tasks, NICER has been pointed at a small set of non-accreting pulsars for more than a million seconds each.  These pulsars have  X-ray emitting ``hot spots" rotating with the star on the stellar surface that are believed to be produced by the impact of high Lorentz factor electrons and positrons on the surface, which are generated as part of the process that produces radio pulsar emission.  NICER records the arrival time of each photon to better than 100 nanosecond accuracy, which is much shorter than the $\sim$few millisecond rotational periods of these pulsars.  The data can therefore be considered as spectra as a function of rotational phase, which is why it is sometimes referred to as time-resolved X-ray spectroscopy.  Although more studies need to be performed, existing work suggests that when rotational phase information as well as spectra are obtained, then if the fit to the data is statistically good, the inferred radius and mass will not be significantly biased.  For example, although the models of the shape and temperature distribution of hot spots cannot be perfectly correct, use of such models will either (1)~produce a statistically poor fit, which then motivates the development of better models prior to radius inference, or (2)~produce a statistically good fit, in which case the inferred radius and mass can provisionally be accepted to be reliable as well as precise.

Like inference from LIGO data, inference of the radius from NICER data proceeds along standard Bayesian lines: given a model for the time-dependent spectra with parameters and associated priors (e.g., the mass, radius, spot shapes, locations, and temperatures, and the observer inclination angle), the NICER team determines the likelihood of the data given the model with specific parameter values, and iterates using a sampler until they obtain the posterior.  Different parameters affect the phase-dependent spectra in ways that can be partially degenerate.  For example, weak modulation could be produced by a very small spot (whose flux might be less than the background flux), or a spot that covers almost the entire star, or a spot nearly centered on the rotational pole, or an observer inclination nearly aligned with the rotational pole.  However, with the hundreds of thousands of counts obtained in NICER observations, these degeneracies can be broken.  

\begin{figure}[ht]
\centering
\includegraphics[width=9.0cm]{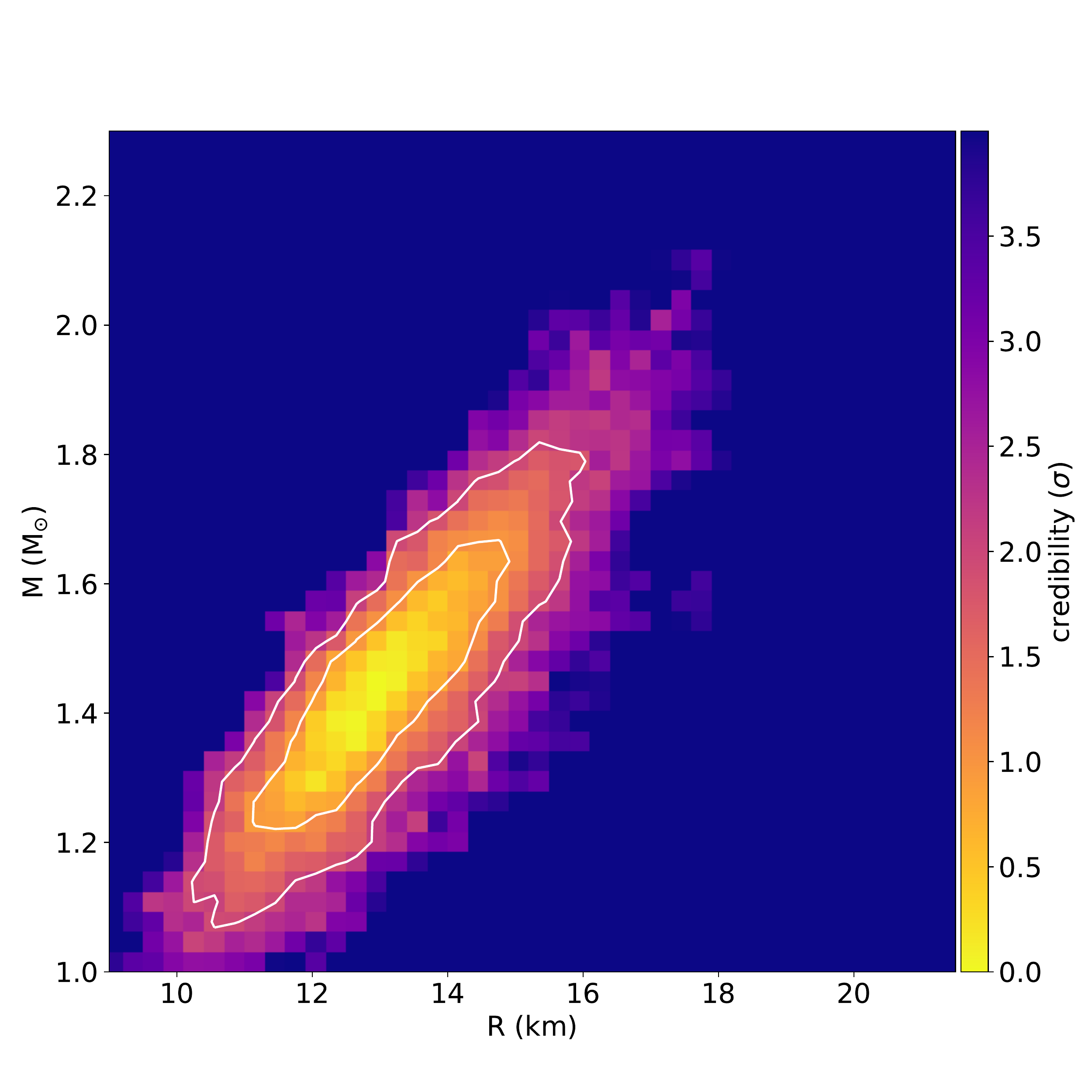}
\includegraphics[width=8.0cm]{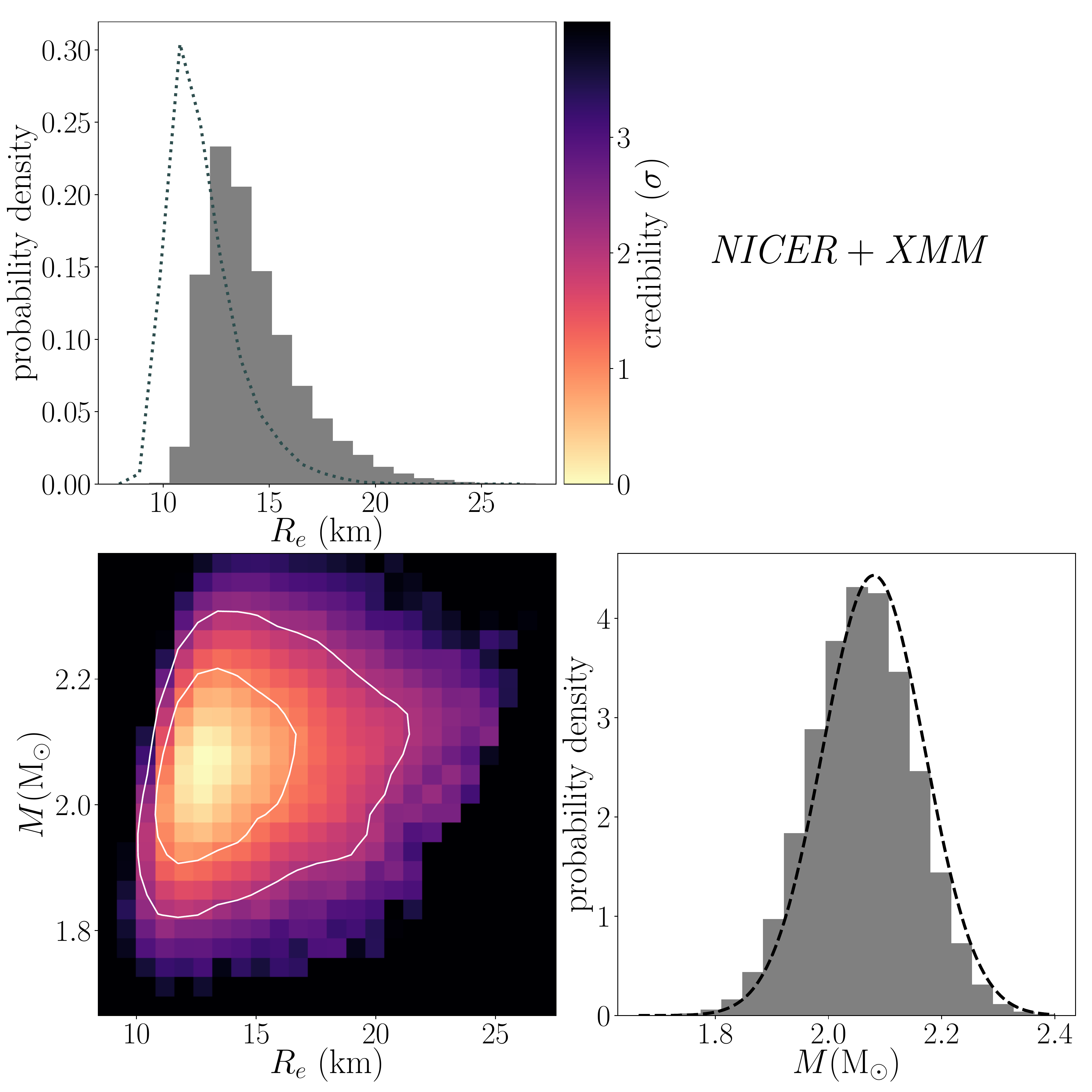}
\caption{Left panel: posterior probability density for the mass and radius of the isolated neutron star PSR~J0030+0451 using only NICER X-ray data (original figure from \cite{2019ApJ...887L..24M}).  Right panel: posterior probability density for the mass and radius of the binary neutron star PSR~J0740+6620, using NICER and XMM-Newton X-ray data as well as Green Bank and CHIME radio data (original figure from \cite{2021arXiv210506979M}).  These results, and the consistent results from the independent analyses of \cite{2019ApJ...887L..21R} and \cite{2021arXiv210506980R}, imply that the radius of a neutron star is roughly constant from $\sim 1.4~M_\odot$ to $\sim 2.1~M_\odot$, and thus that matter in the core of neutron stars has relatively high pressure.  
}
\label{fig:NICERmr}
\end{figure}

At present, two independent groups within the NICER team have inferred the mass and radius of two pulsars: PSR~J0300+0451 and PSR~J0740+6620. The first one is an isolated pulsar (which, being isolated, does not have an independently measured mass) that spins at a frequency of 205.53~Hz. For PSR~J0030+0451 the teams reported a mass of $\approx 1.4~M_\odot$ and 68.3\% credible regions on the radius of $12.0-14.3$~km (Miller et al. 2019~\cite{2019ApJ...887L..24M}; see the left panel of Figure~\ref{fig:NICERmr}) and $11.5-13.9$~km (Riley et al. 2019~\cite{2019ApJ...887L..21R}).  The teams also found very similar hot spot locations and shapes.  These radii are compatible with the 90\% credibility radius upper bound of $\sim 13.5$~km found from GW170817, which contained two neutron stars which also had masses $\sim 1.3-1.4~M_\odot$.

The second pulsar, PSR~J0740+6620, spins at 346.53~Hz and is $\sim 20\times$ fainter in X-rays than PSR~J0300+0451, but it is in a binary system, so one possesses independent measurements of the mass and observer inclination angle from radio observations. The Green Bank and CHIME radio telescopes have inferred a mass of $M=2.08\pm 0.07~M_\odot$ and an observer inclination to the binary orbital axis of $\theta_{\rm obs}\approx 87.5^\circ$. The inclusion of this independent information is crucial to obtain a good measurement of the neutron star radius. In addition, XMM-Newton data for this pulsar produced a more precise estimate of the pulsar X-ray flux than was possible using NICER alone, given the low flux of the pulsar and the comparatively high X-ray background in NICER observations. Combining the NICER, the XMM-Newton, and the radio data, the two teams reported 68.3\% credible regions on the radius of $12.2-16.3$~km (Miller et al. 2021~\cite{2021arXiv210506979M}; see the right panel of Figure~\ref{fig:NICERmr}) and $11.4-13.7$~km (Riley et al. 2021~\cite{2021arXiv210506980R}). The difference in the credible regions is primarily due to the use of different statistical samplers and different assumptions about the cross-calibration between NICER and XMM-Newton.

The net result of the NICER measurements is that the radii of $\sim 1.4~M_\odot$ and $\sim 2.1~M_\odot$ neutron stars are not very different from each other (indeed, they are consistent with being the same), with both being on the order of $12-14$~km.  This implies a relatively hard equation of state, with a maximum mass sufficiently above the $\sim 2.1~M_\odot$ mass of PSR~J0740+6620 that the radius has not yet bent toward smaller values (which is characteristic of the mass-radius curve near the maximum mass). The tidal deformability measurement for GW170817 eliminates the hardest equations of state, so together NICER and gravitational wave measurements have significantly narrowed the plausible list of candidates for the high-density equation of state.

\section{The beauty of future dreams}

In the last 6 years, we have gone from living in a metaphorical data desert, with only a handful of gravitational wave observations, to living in a metaphorical data forest, with over 50 observations and counting. Clearly, gravitational wave astrophysics is here to stay, and we have only but began to explore this forest. The fourth observing (O4) run of the LIGO and Virgo detectors, which is expected to reach design sensitivity, is scheduled to start in 2022, and the fifth observing (O5) run, which is expected to reach even better sensitivities, will start a few years later. The observing range for O4 will be about 50\%--90\% larger than that of O3, and the range for O5 is expected to be three times larger than that of O3. Since at the low redshifts relevant for double neutron star observations the accessible volume scales with the cube of the observing range, one can expect many, many more observations of binary black holes, binary neutron stars and mixed binaries, perhaps even in the hundreds by the time of O5 in just 1 year of data. By the mid-2030s, third generation ground-based detectors such as the Einstein Telescope and the Cosmic explorer will improve sensitivity by an additional factor of $10-20$.

What will the advent of such an increase in sensitivity do for us? As we expect to detect many more events, we in particular expect more binary neutron star observations. The low signal-to-noise ratio binary neutron star events may be too far away to lead to an electromagnetic counterpart, but if some of these mergers occur close to the Milky Way (say at tens of Mpc), then there is a good chance of a GW170817 repeat. This time, however, since the gravitational wave sensitivity will be significantly larger, the signal-to-noise ratio for an event at 40Mpc could be three times larger than that of GW170817, and thus, close to 100. Such a high signal-to-noise ratio gravitational wave event would inaugurate the era of \textit{precision} gravitational wave nuclear astrophysics, since it would allow for a much more accurate measurement of the tidal deformabilities, and thus of the radius of neutron stars. In fact, for such a loud event, not only could one extract nuclear physics information from the inspiral phase of the event, but one may also be able to extract information from the merger and post-merger itself, which may be useful to probe e.g. the presence of a quark matter core inside hybrid stars~\cite{Bauswein:2020ggy}.

Meanwhile, NICER observations will continue, and in particular, more and more data will be accumulated of the (soon to be) three pulsars that have already been observed. In fact, in 2022 the NICER team is expected to publish the measurement of the radius of the third pulsar, and to update the estimated radii of PSR J0030+0451 and PSR J0740+6620 using new data and a better characterization of the instrument.  These measurements will improve the precision of our knowledge of the sizes of neutron stars over a factor of 1.5 in mass, which will therefore yield additional valuable information about the equation of state of matter beyond nuclear density and the existence of quark matter~\cite{Blaschke:2020vuy} inside neutron star cores.  Beyond NICER there are numerous planned and proposed facilities and missions that will extend our reach greatly: these include radio observatories (the Square Kilometer Array and the next generation Very Large Array) and numerous X-ray missions (such as Athena, the enhanced X-ray Timing and Polarimetry mission [eXTP], and STROBE$-$X), which can probe the masses, radii, moments of inertia, and cooling properties of neutron stars.

Whatever the future may hold, what is clear is that the combination of information from electromagnetic and gravitational wave observations is revealing the states of matter that are realized in the cores of neutron stars, at wonderfully large pressures and densities. In fact, it is not unreasonable to wager that within the next 10 years, the equation of state of matter at a few times nuclear saturation density will be, for the first time, constrained to better than 10\% in the low-temperature neutron star region of the quantum chromodynamics phase space. The challenge now and in the near future will be to find ever more creative ways to connect these observations to fundamental nuclear physics. The future, is thus, beautifully exciting. 

\section*{Acknowledgements}
We thank Katerina Chatziioannou and Jaki Noronha-Hostler for carefully reading the manuscript and giving us valuable comments.
N.~Y.~ acknowledges support from NSF grant AST Award No.~2009268 and the Simons Foundation.
M.C.M acknowledges support from NASA ADAP grant 80NSSC21K0649.  He performed part of his work on this paper at the Aspen Center for Physics, which is supported by National Science Foundation grant PHY-1607611.
K.Y. acknowledges support from NSF Grant PHY-1806776, NASA Grant 80NSSC20K0523, a Sloan Foundation Research Fellowship and the Owens Family Foundation.

\section*{Author contributions}
The authors contributed equally to all aspects of the article. 

\section*{Competing interests}
The authors declare no competing interests. 

\section*{Publisher’s note}
Springer Nature remains neutral with regard to jurisdictional claims in published maps and institutional affiliations.

\bibliography{sample}

\begin{thebibliography}{10}
\urlstyle{rm}
\expandafter\ifx\csname url\endcsname\relax
  \def\url#1{\texttt{#1}}\fi
\expandafter\ifx\csname urlprefix\endcsname\relax\def\urlprefix{URL }\fi
\expandafter\ifx\csname doiprefix\endcsname\relax\def\doiprefix{DOI: }\fi
\providecommand{\bibinfo}[2]{#2}
\providecommand{\eprint}[2][]{\url{#2}}

\bibitem{Shapiro:1983du}
\bibinfo{author}{Shapiro, S.~L.} \& \bibinfo{author}{Teukolsky, S.~A.}
\newblock \emph{\bibinfo{title}{{Black holes, white dwarfs, and neutron stars:
  The physics of compact objects}}} (\bibinfo{year}{1983}).

\bibitem{Kaspi:2017fwg}
\bibinfo{author}{Kaspi, V.~M.} \& \bibinfo{author}{Beloborodov, A.}
\newblock \bibinfo{journal}{\bibinfo{title}{{Magnetars}}}.
\newblock {\emph{\JournalTitle{Ann. Rev. Astron. Astrophys.}}}
  \textbf{\bibinfo{volume}{55}}, \bibinfo{pages}{261--301},
  \doiprefix\url{10.1146/annurev-astro-081915-023329} (\bibinfo{year}{2017}).
\newblock \eprint{1703.00068}.

\bibitem{doi:10.1126/science.238.4828.761}
\bibinfo{author}{Rawley, L.~A.}, \bibinfo{author}{Taylor, J.~H.},
  \bibinfo{author}{Davis, M.~M.} \& \bibinfo{author}{Allan, D.~W.}
\newblock \bibinfo{journal}{\bibinfo{title}{Millisecond pulsar psr 1937+21: A
  highly stable clock}}.
\newblock {\emph{\JournalTitle{Science}}} \textbf{\bibinfo{volume}{238}},
  \bibinfo{pages}{761--765}, \doiprefix\url{10.1126/science.238.4828.761}
  (\bibinfo{year}{1987}).
\newblock
  \eprint{https://www.science.org/doi/pdf/10.1126/science.238.4828.761}.

\bibitem{Baym:1971pw}
\bibinfo{author}{Baym, G.}, \bibinfo{author}{Pethick, C.} \&
  \bibinfo{author}{Sutherland, P.}
\newblock \bibinfo{journal}{\bibinfo{title}{{The Ground state of matter at high
  densities: Equation of state and stellar models}}}.
\newblock {\emph{\JournalTitle{Astrophys. J.}}} \textbf{\bibinfo{volume}{170}},
  \bibinfo{pages}{299--317}, \doiprefix\url{10.1086/151216}
  (\bibinfo{year}{1971}).

\bibitem{Ambartsumyan}
\bibinfo{journal}{\bibinfo{author}{Ambartsumyan, G.~S., V.~A.~and~Saakyan}}.
\newblock {\emph{\JournalTitle{Sov. Astron.}}} \textbf{\bibinfo{volume}{4}},
  \bibinfo{pages}{187} (\bibinfo{year}{1960}).

\bibitem{Chatterjee:2015pua}
\bibinfo{author}{Chatterjee, D.} \& \bibinfo{author}{Vida\~na, I.}
\newblock \bibinfo{journal}{\bibinfo{title}{{Do hyperons exist in the interior
  of neutron stars?}}}
\newblock {\emph{\JournalTitle{Eur. Phys. J. A}}}
  \textbf{\bibinfo{volume}{52}}, \bibinfo{pages}{29},
  \doiprefix\url{10.1140/epja/i2016-16029-x} (\bibinfo{year}{2016}).
\newblock \eprint{1510.06306}.

\bibitem{Ivanenko:1965dg}
\bibinfo{author}{Ivanenko, D.~D.} \& \bibinfo{author}{Kurdgelaidze, D.~F.}
\newblock \bibinfo{journal}{\bibinfo{title}{{Hypothesis concerning quark
  stars}}}.
\newblock {\emph{\JournalTitle{Astrophysics}}} \textbf{\bibinfo{volume}{1}},
  \bibinfo{pages}{251--252}, \doiprefix\url{10.1007/BF01042830}
  (\bibinfo{year}{1965}).

\bibitem{lattimer_prakash2001}
\bibinfo{author}{Lattimer, J.~M.} \& \bibinfo{author}{Prakash, M.}
\newblock \bibinfo{journal}{\bibinfo{title}{Neutron star structure and the
  equation of state}}.
\newblock {\emph{\JournalTitle{Astrophys. J.}}} \textbf{\bibinfo{volume}{550}},
  \bibinfo{pages}{426} (\bibinfo{year}{2001}).

\bibitem{Most:2021zvc}
\bibinfo{author}{Most, E.~R.} \emph{et~al.}
\newblock \bibinfo{journal}{\bibinfo{title}{{Projecting the likely importance
  of weak-interaction-driven bulk viscosity in neutron star mergers}}}.
\newblock {\emph{\JournalTitle{{submitted to MNRAS}}}}  (\bibinfo{year}{2021}).
\newblock \eprint{2107.05094}.

\bibitem{Troyer:2004ge}
\bibinfo{author}{Troyer, M.} \& \bibinfo{author}{Wiese, U.-J.}
\newblock \bibinfo{journal}{\bibinfo{title}{{Computational complexity and
  fundamental limitations to fermionic quantum Monte Carlo simulations}}}.
\newblock {\emph{\JournalTitle{Phys. Rev. Lett.}}}
  \textbf{\bibinfo{volume}{94}}, \bibinfo{pages}{170201},
  \doiprefix\url{10.1103/PhysRevLett.94.170201} (\bibinfo{year}{2005}).
\newblock \eprint{cond-mat/0408370}.

\bibitem{Aoki:2006we}
\bibinfo{author}{Aoki, Y.}, \bibinfo{author}{Endrodi, G.},
  \bibinfo{author}{Fodor, Z.}, \bibinfo{author}{Katz, S.~D.} \&
  \bibinfo{author}{Szabo, K.~K.}
\newblock \bibinfo{journal}{\bibinfo{title}{{The Order of the quantum
  chromodynamics transition predicted by the standard model of particle
  physics}}}.
\newblock {\emph{\JournalTitle{Nature}}} \textbf{\bibinfo{volume}{443}},
  \bibinfo{pages}{675--678}, \doiprefix\url{10.1038/nature05120}
  (\bibinfo{year}{2006}).
\newblock \eprint{hep-lat/0611014}.

\bibitem{Baym:2017whm}
\bibinfo{author}{Baym, G.} \emph{et~al.}
\newblock \bibinfo{journal}{\bibinfo{title}{{From hadrons to quarks in neutron
  stars: a review}}}.
\newblock {\emph{\JournalTitle{Rept. Prog. Phys.}}}
  \textbf{\bibinfo{volume}{81}}, \bibinfo{pages}{056902},
  \doiprefix\url{10.1088/1361-6633/aaae14} (\bibinfo{year}{2018}).
\newblock \eprint{1707.04966}.

\bibitem{Nambu:1961tp}
\bibinfo{author}{Nambu, Y.} \& \bibinfo{author}{Jona-Lasinio, G.}
\newblock \bibinfo{journal}{\bibinfo{title}{{Dynamical Model of Elementary
  Particles Based on an Analogy with Superconductivity. 1.}}}
\newblock {\emph{\JournalTitle{Phys. Rev.}}} \textbf{\bibinfo{volume}{122}},
  \bibinfo{pages}{345--358}, \doiprefix\url{10.1103/PhysRev.122.345}
  (\bibinfo{year}{1961}).

\bibitem{Alford:1997zt}
\bibinfo{author}{Alford, M.~G.}, \bibinfo{author}{Rajagopal, K.} \&
  \bibinfo{author}{Wilczek, F.}
\newblock \bibinfo{journal}{\bibinfo{title}{{QCD at finite baryon density:
  Nucleon droplets and color superconductivity}}}.
\newblock {\emph{\JournalTitle{Phys. Lett. B}}} \textbf{\bibinfo{volume}{422}},
  \bibinfo{pages}{247--256}, \doiprefix\url{10.1016/S0370-2693(98)00051-3}
  (\bibinfo{year}{1998}).
\newblock \eprint{hep-ph/9711395}.

\bibitem{Dexheimer:2009hi}
\bibinfo{author}{Dexheimer, V.~A.} \& \bibinfo{author}{Schramm, S.}
\newblock \bibinfo{journal}{\bibinfo{title}{{A Novel Approach to Model Hybrid
  Stars}}}.
\newblock {\emph{\JournalTitle{Phys. Rev. C}}} \textbf{\bibinfo{volume}{81}},
  \bibinfo{pages}{045201}, \doiprefix\url{10.1103/PhysRevC.81.045201}
  (\bibinfo{year}{2010}).
\newblock \eprint{0901.1748}.

\bibitem{Tews:2012fj}
\bibinfo{author}{Tews, I.}, \bibinfo{author}{Kr\"uger, T.},
  \bibinfo{author}{Hebeler, K.} \& \bibinfo{author}{Schwenk, A.}
\newblock \bibinfo{journal}{\bibinfo{title}{{Neutron matter at
  next-to-next-to-next-to-leading order in chiral effective field theory}}}.
\newblock {\emph{\JournalTitle{Phys. Rev. Lett.}}}
  \textbf{\bibinfo{volume}{110}}, \bibinfo{pages}{032504},
  \doiprefix\url{10.1103/PhysRevLett.110.032504} (\bibinfo{year}{2013}).
\newblock \eprint{1206.0025}.

\bibitem{Read:2008iy}
\bibinfo{author}{Read, J.~S.}, \bibinfo{author}{Lackey, B.~D.},
  \bibinfo{author}{Owen, B.~J.} \& \bibinfo{author}{Friedman, J.~L.}
\newblock \bibinfo{journal}{\bibinfo{title}{{Constraints on a
  phenomenologically parameterized neutron-star equation of state}}}.
\newblock {\emph{\JournalTitle{Phys. Rev. D}}} \textbf{\bibinfo{volume}{79}},
  \bibinfo{pages}{124032}, \doiprefix\url{10.1103/PhysRevD.79.124032}
  (\bibinfo{year}{2009}).
\newblock \eprint{0812.2163}.

\bibitem{Annala:2017llu}
\bibinfo{author}{Annala, E.}, \bibinfo{author}{Gorda, T.},
  \bibinfo{author}{Kurkela, A.} \& \bibinfo{author}{Vuorinen, A.}
\newblock \bibinfo{journal}{\bibinfo{title}{{Gravitational-wave constraints on
  the neutron-star-matter Equation of State}}}.
\newblock {\emph{\JournalTitle{Phys. Rev. Lett.}}}
  \textbf{\bibinfo{volume}{120}}, \bibinfo{pages}{172703},
  \doiprefix\url{10.1103/PhysRevLett.120.172703} (\bibinfo{year}{2018}).
\newblock \eprint{1711.02644}.

\bibitem{Alford:2013aca}
\bibinfo{author}{Alford, M.~G.}, \bibinfo{author}{Han, S.} \&
  \bibinfo{author}{Prakash, M.}
\newblock \bibinfo{journal}{\bibinfo{title}{{Generic conditions for stable
  hybrid stars}}}.
\newblock {\emph{\JournalTitle{Phys. Rev. D}}} \textbf{\bibinfo{volume}{88}},
  \bibinfo{pages}{083013}, \doiprefix\url{10.1103/PhysRevD.88.083013}
  (\bibinfo{year}{2013}).
\newblock \eprint{1302.4732}.

\bibitem{Haque:2014rua}
\bibinfo{author}{Haque, N.} \emph{et~al.}
\newblock \bibinfo{journal}{\bibinfo{title}{{Three-loop HTLpt thermodynamics at
  finite temperature and chemical potential}}}.
\newblock {\emph{\JournalTitle{JHEP}}} \textbf{\bibinfo{volume}{05}},
  \bibinfo{pages}{027}, \doiprefix\url{10.1007/JHEP05(2014)027}
  (\bibinfo{year}{2014}).
\newblock \eprint{1402.6907}.

\bibitem{Glendenning:1998ag}
\bibinfo{author}{Glendenning, N.~K.} \& \bibinfo{author}{Kettner, C.}
\newblock \bibinfo{journal}{\bibinfo{title}{{Nonidentical neutron star
  twins}}}.
\newblock {\emph{\JournalTitle{Astron. Astrophys.}}}
  \textbf{\bibinfo{volume}{353}}, \bibinfo{pages}{L9} (\bibinfo{year}{2000}).
\newblock \eprint{astro-ph/9807155}.

\bibitem{Page:2004fy}
\bibinfo{author}{Page, D.}, \bibinfo{author}{Lattimer, J.~M.},
  \bibinfo{author}{Prakash, M.} \& \bibinfo{author}{Steiner, A.~W.}
\newblock \bibinfo{journal}{\bibinfo{title}{{Minimal cooling of neutron stars:
  A New paradigm}}}.
\newblock {\emph{\JournalTitle{Astrophys. J. Suppl.}}}
  \textbf{\bibinfo{volume}{155}}, \bibinfo{pages}{623--650},
  \doiprefix\url{10.1086/424844} (\bibinfo{year}{2004}).
\newblock \eprint{astro-ph/0403657}.

\bibitem{Blaschke:2004vq}
\bibinfo{author}{Blaschke, D.}, \bibinfo{author}{Grigorian, H.} \&
  \bibinfo{author}{Voskresensky, D.~N.}
\newblock \bibinfo{journal}{\bibinfo{title}{{Cooling of neutron stars: Hadronic
  model}}}.
\newblock {\emph{\JournalTitle{Astron. Astrophys.}}}
  \textbf{\bibinfo{volume}{424}}, \bibinfo{pages}{979--992},
  \doiprefix\url{10.1051/0004-6361:20040404} (\bibinfo{year}{2004}).
\newblock \eprint{astro-ph/0403170}.

\bibitem{Piekarewicz:2014lba}
\bibinfo{author}{Piekarewicz, J.}, \bibinfo{author}{Fattoyev, F.~J.} \&
  \bibinfo{author}{Horowitz, C.~J.}
\newblock \bibinfo{journal}{\bibinfo{title}{{Pulsar Glitches: The Crust may be
  Enough}}}.
\newblock {\emph{\JournalTitle{Phys. Rev. C}}} \textbf{\bibinfo{volume}{90}},
  \bibinfo{pages}{015803}, \doiprefix\url{10.1103/PhysRevC.90.015803}
  (\bibinfo{year}{2014}).
\newblock \eprint{1404.2660}.

\bibitem{Haskell:2015jra}
\bibinfo{author}{Haskell, B.} \& \bibinfo{author}{Melatos, A.}
\newblock \bibinfo{journal}{\bibinfo{title}{{Models of Pulsar Glitches}}}.
\newblock {\emph{\JournalTitle{Int. J. Mod. Phys. D}}}
  \textbf{\bibinfo{volume}{24}}, \bibinfo{pages}{1530008},
  \doiprefix\url{10.1142/S0218271815300086} (\bibinfo{year}{2015}).
\newblock \eprint{1502.07062}.

\bibitem{2018arXiv180404952F}
\bibinfo{author}{{Fattoyev}, F.~J.}, \bibinfo{author}{{Horowitz}, C.~J.} \&
  \bibinfo{author}{{Lu}, H.}
\newblock \bibinfo{journal}{\bibinfo{title}{{Crust breaking and the limiting
  rotational frequency of neutron stars}}}.
\newblock {\emph{\JournalTitle{arXiv e-prints}}}
  \bibinfo{pages}{arXiv:1804.04952} (\bibinfo{year}{2018}).
\newblock \eprint{1804.04952}.

\bibitem{2021arXiv210403137H}
\bibinfo{author}{{Haskell}, B.} \& \bibinfo{author}{{Schwenzer}, K.}
\newblock \bibinfo{journal}{\bibinfo{title}{{Gravitational waves from isolated
  neutron stars}}}.
\newblock {\emph{\JournalTitle{arXiv e-prints}}}
  \bibinfo{pages}{arXiv:2104.03137} (\bibinfo{year}{2021}).
\newblock \eprint{2104.03137}.

\bibitem{Hulse:1974eb}
\bibinfo{author}{Hulse, R.~A.} \& \bibinfo{author}{Taylor, J.~H.}
\newblock \bibinfo{journal}{\bibinfo{title}{{Discovery of a pulsar in a binary
  system}}}.
\newblock {\emph{\JournalTitle{Astrophys. J. Lett.}}}
  \textbf{\bibinfo{volume}{195}}, \bibinfo{pages}{L51--L53},
  \doiprefix\url{10.1086/181708} (\bibinfo{year}{1975}).

\bibitem{Cromartie:2019kug}
\bibinfo{author}{Cromartie, H.~T.} \emph{et~al.}
\newblock \bibinfo{journal}{\bibinfo{title}{{Relativistic Shapiro delay
  measurements of an extremely massive millisecond pulsar}}}.
\newblock {\emph{\JournalTitle{Nature Astron.}}} \textbf{\bibinfo{volume}{4}},
  \bibinfo{pages}{72--76}, \doiprefix\url{10.1038/s41550-019-0880-2}
  (\bibinfo{year}{2019}).
\newblock \eprint{1904.06759}.

\bibitem{1.97NS}
\bibinfo{author}{{Demorest}, P.~B.}, \bibinfo{author}{{Pennucci}, T.},
  \bibinfo{author}{{Ransom}, S.~M.}, \bibinfo{author}{{Roberts}, M.~S.~E.} \&
  \bibinfo{author}{{Hessels}, J.~W.~T.}
\newblock \bibinfo{journal}{\bibinfo{title}{{A two-solar-mass neutron star
  measured using Shapiro delay}}}.
\newblock {\emph{\JournalTitle{Nature}}} \textbf{\bibinfo{volume}{467}},
  \bibinfo{pages}{1081--1083}, \doiprefix\url{10.1038/nature09466}
  (\bibinfo{year}{2010}).
\newblock \eprint{1010.5788}.

\bibitem{2.01NS}
\bibinfo{author}{Antoniadis, J.} \emph{et~al.}
\newblock \bibinfo{journal}{\bibinfo{title}{{A Massive Pulsar in a Compact
  Relativistic Binary}}}.
\newblock {\emph{\JournalTitle{Science}}} \textbf{\bibinfo{volume}{340}},
  \bibinfo{pages}{6131}, \doiprefix\url{10.1126/science.1233232}
  (\bibinfo{year}{2013}).
\newblock \eprint{1304.6875}.

\bibitem{Abbott:2016blz}
\bibinfo{author}{Abbott, B.~P.} \emph{et~al.}
\newblock \bibinfo{journal}{\bibinfo{title}{{Observation of Gravitational Waves
  from a Binary Black Hole Merger}}}.
\newblock {\emph{\JournalTitle{Phys. Rev. Lett.}}}
  \textbf{\bibinfo{volume}{116}}, \bibinfo{pages}{061102},
  \doiprefix\url{10.1103/PhysRevLett.116.061102} (\bibinfo{year}{2016}).

\bibitem{LIGOScientific:2020ibl}
\bibinfo{author}{Abbott, R.} \emph{et~al.}
\newblock \bibinfo{journal}{\bibinfo{title}{{GWTC-2: Compact Binary
  Coalescences Observed by LIGO and Virgo During the First Half of the Third
  Observing Run}}}.
\newblock {\emph{\JournalTitle{Phys. Rev. X}}} \textbf{\bibinfo{volume}{11}},
  \bibinfo{pages}{021053}, \doiprefix\url{10.1103/PhysRevX.11.021053}
  (\bibinfo{year}{2021}).
\newblock \eprint{2010.14527}.

\bibitem{TheLIGOScientific:2017qsa}
\bibinfo{author}{Abbott, B.~P.} \emph{et~al.}
\newblock \bibinfo{journal}{\bibinfo{title}{{GW170817: Observation of
  Gravitational Waves from a Binary Neutron Star Inspiral}}}.
\newblock {\emph{\JournalTitle{Phys. Rev. Lett.}}}
  \textbf{\bibinfo{volume}{119}}, \bibinfo{pages}{161101},
  \doiprefix\url{10.1103/PhysRevLett.119.161101} (\bibinfo{year}{2017}).

\bibitem{flanagan-hinderer-love}
\bibinfo{author}{Flanagan, E.~E.} \& \bibinfo{author}{Hinderer, T.}
\newblock \bibinfo{journal}{\bibinfo{title}{{Constraining neutron star tidal
  Love numbers with gravitational wave detectors}}}.
\newblock {\emph{\JournalTitle{Phys.Rev.}}} \textbf{\bibinfo{volume}{D77}},
  \bibinfo{pages}{021502}, \doiprefix\url{10.1103/PhysRevD.77.021502}
  (\bibinfo{year}{2008}).

\bibitem{hinderer-lackey-lang-read}
\bibinfo{author}{Hinderer, T.}, \bibinfo{author}{Lackey, B.~D.},
  \bibinfo{author}{Lang, R.~N.} \& \bibinfo{author}{Read, J.~S.}
\newblock \bibinfo{journal}{\bibinfo{title}{{Tidal deformability of neutron
  stars with realistic equations of state and their gravitational wave
  signatures in binary inspiral}}}.
\newblock {\emph{\JournalTitle{Phys.Rev.}}} \textbf{\bibinfo{volume}{D81}},
  \bibinfo{pages}{123016}, \doiprefix\url{10.1103/PhysRevD.81.123016}
  (\bibinfo{year}{2010}).

\bibitem{Carson:2020gwh}
\bibinfo{author}{{Carson}, Z.}
\newblock \bibinfo{journal}{\bibinfo{title}{{Probing Fundamental Physics with
  Gravitational Waves}}}.
\newblock {\emph{\JournalTitle{University of Virginia (Ph.D thesis)}}}
  (\bibinfo{year}{2020}).
\newblock \eprint{2010.04745}.

\bibitem{LIGOScientific:2018cki}
\bibinfo{author}{Abbott, B.~P.} \emph{et~al.}
\newblock \bibinfo{journal}{\bibinfo{title}{{GW170817: Measurements of neutron
  star radii and equation of state}}}.
\newblock {\emph{\JournalTitle{Phys. Rev. Lett.}}}
  \textbf{\bibinfo{volume}{121}}, \bibinfo{pages}{161101},
  \doiprefix\url{10.1103/PhysRevLett.121.161101} (\bibinfo{year}{2018}).
\newblock \eprint{1805.11581}.

\bibitem{Chatziioannou:2020pqz}
\bibinfo{author}{Chatziioannou, K.}
\newblock \bibinfo{journal}{\bibinfo{title}{{Neutron star tidal deformability
  and equation of state constraints}}}.
\newblock {\emph{\JournalTitle{Gen. Rel. Grav.}}}
  \textbf{\bibinfo{volume}{52}}, \bibinfo{pages}{109},
  \doiprefix\url{10.1007/s10714-020-02754-3} (\bibinfo{year}{2020}).
\newblock \eprint{2006.03168}.

\bibitem{Lindblom:2010bb}
\bibinfo{author}{Lindblom, L.}
\newblock \bibinfo{journal}{\bibinfo{title}{{Spectral Representations of
  Neutron-Star Equations of State}}}.
\newblock {\emph{\JournalTitle{Phys. Rev. D}}} \textbf{\bibinfo{volume}{82}},
  \bibinfo{pages}{103011}, \doiprefix\url{10.1103/PhysRevD.82.103011}
  (\bibinfo{year}{2010}).
\newblock \eprint{1009.0738}.

\bibitem{Greif:2018njt}
\bibinfo{author}{Greif, S.~K.}, \bibinfo{author}{Raaijmakers, G.},
  \bibinfo{author}{Hebeler, K.}, \bibinfo{author}{Schwenk, A.} \&
  \bibinfo{author}{Watts, A.~L.}
\newblock \bibinfo{journal}{\bibinfo{title}{{Equation of state sensitivities
  when inferring neutron star and dense matter properties}}}.
\newblock {\emph{\JournalTitle{Mon. Not. Roy. Astron. Soc.}}}
  \textbf{\bibinfo{volume}{485}}, \bibinfo{pages}{5363--5376},
  \doiprefix\url{10.1093/mnras/stz654} (\bibinfo{year}{2019}).
\newblock \eprint{1812.08188}.

\bibitem{Tews:2018kmu}
\bibinfo{author}{Tews, I.}, \bibinfo{author}{Carlson, J.},
  \bibinfo{author}{Gandolfi, S.} \& \bibinfo{author}{Reddy, S.}
\newblock \bibinfo{journal}{\bibinfo{title}{{Constraining the speed of sound
  inside neutron stars with chiral effective field theory interactions and
  observations}}}.
\newblock {\emph{\JournalTitle{Astrophys. J.}}} \textbf{\bibinfo{volume}{860}},
  \bibinfo{pages}{149}, \doiprefix\url{10.3847/1538-4357/aac267}
  (\bibinfo{year}{2018}).
\newblock \eprint{1801.01923}.

\bibitem{Yagi:2016bkt}
\bibinfo{author}{Yagi, K.} \& \bibinfo{author}{Yunes, N.}
\newblock \bibinfo{journal}{\bibinfo{title}{{Approximate Universal Relations
  for Neutron Stars and Quark Stars}}}.
\newblock {\emph{\JournalTitle{Phys. Rept.}}} \textbf{\bibinfo{volume}{681}},
  \bibinfo{pages}{1--72}, \doiprefix\url{10.1016/j.physrep.2017.03.002}
  (\bibinfo{year}{2017}).
\newblock \eprint{1608.02582}.

\bibitem{Yagi:2013bca}
\bibinfo{author}{Yagi, K.} \& \bibinfo{author}{Yunes, N.}
\newblock \bibinfo{journal}{\bibinfo{title}{{I-Love-Q}}}.
\newblock {\emph{\JournalTitle{Science}}} \textbf{\bibinfo{volume}{341}},
  \bibinfo{pages}{365--368}, \doiprefix\url{10.1126/science.1236462}
  (\bibinfo{year}{2013}).
\newblock \eprint{1302.4499}.

\bibitem{Yagi:2013awa}
\bibinfo{author}{Yagi, K.} \& \bibinfo{author}{Yunes, N.}
\newblock \bibinfo{journal}{\bibinfo{title}{{I-Love-Q Relations in Neutron
  Stars and their Applications to Astrophysics, Gravitational Waves and
  Fundamental Physics}}}.
\newblock {\emph{\JournalTitle{Phys. Rev. D}}} \textbf{\bibinfo{volume}{88}},
  \bibinfo{pages}{023009}, \doiprefix\url{10.1103/PhysRevD.88.023009}
  (\bibinfo{year}{2013}).
\newblock \eprint{1303.1528}.

\bibitem{Yagi:2015pkc}
\bibinfo{author}{Yagi, K.} \& \bibinfo{author}{Yunes, N.}
\newblock \bibinfo{journal}{\bibinfo{title}{{Binary Love Relations}}}.
\newblock {\emph{\JournalTitle{Class. Quant. Grav.}}}
  \textbf{\bibinfo{volume}{33}}, \bibinfo{pages}{13LT01},
  \doiprefix\url{10.1088/0264-9381/33/13/13LT01} (\bibinfo{year}{2016}).
\newblock \eprint{1512.02639}.

\bibitem{Yagi:2016qmr}
\bibinfo{author}{Yagi, K.} \& \bibinfo{author}{Yunes, N.}
\newblock \bibinfo{journal}{\bibinfo{title}{{Approximate Universal Relations
  among Tidal Parameters for Neutron Star Binaries}}}.
\newblock {\emph{\JournalTitle{Class. Quant. Grav.}}}
  \textbf{\bibinfo{volume}{34}}, \bibinfo{pages}{015006},
  \doiprefix\url{10.1088/1361-6382/34/1/015006} (\bibinfo{year}{2017}).
\newblock \eprint{1608.06187}.

\bibitem{De:2018uhw}
\bibinfo{author}{De, S.} \emph{et~al.}
\newblock \bibinfo{journal}{\bibinfo{title}{{Tidal Deformabilities and Radii of
  Neutron Stars from the Observation of GW170817}}}.
\newblock {\emph{\JournalTitle{Phys. Rev. Lett.}}}
  \textbf{\bibinfo{volume}{121}}, \bibinfo{pages}{091102},
  \doiprefix\url{10.1103/PhysRevLett.121.091102} (\bibinfo{year}{2018}).
\newblock \bibinfo{note}{[Erratum: Phys.Rev.Lett. 121, 259902 (2018)]},
  \eprint{1804.08583}.

\bibitem{Maselli:2013mva}
\bibinfo{author}{Maselli, A.}, \bibinfo{author}{Cardoso, V.},
  \bibinfo{author}{Ferrari, V.}, \bibinfo{author}{Gualtieri, L.} \&
  \bibinfo{author}{Pani, P.}
\newblock \bibinfo{journal}{\bibinfo{title}{{Equation-of-state-independent
  relations in neutron stars}}}.
\newblock {\emph{\JournalTitle{Phys. Rev. D}}} \textbf{\bibinfo{volume}{88}},
  \bibinfo{pages}{023007}, \doiprefix\url{10.1103/PhysRevD.88.023007}
  (\bibinfo{year}{2013}).
\newblock \eprint{1304.2052}.

\bibitem{Chatziioannou:2018vzf}
\bibinfo{author}{Chatziioannou, K.}, \bibinfo{author}{Haster, C.-J.} \&
  \bibinfo{author}{Zimmerman, A.}
\newblock \bibinfo{journal}{\bibinfo{title}{{Measuring the neutron star tidal
  deformability with equation-of-state-independent relations and gravitational
  waves}}}.
\newblock {\emph{\JournalTitle{Phys. Rev. D}}} \textbf{\bibinfo{volume}{97}},
  \bibinfo{pages}{104036}, \doiprefix\url{10.1103/PhysRevD.97.104036}
  (\bibinfo{year}{2018}).
\newblock \eprint{1804.03221}.

\bibitem{hinderer-love}
\bibinfo{author}{Hinderer, T.}
\newblock \bibinfo{journal}{\bibinfo{title}{{Tidal Love numbers of neutron
  stars}}}.
\newblock {\emph{\JournalTitle{Astrophys.J.}}} \textbf{\bibinfo{volume}{677}},
  \bibinfo{pages}{1216--1220}, \doiprefix\url{10.1086/533487}
  (\bibinfo{year}{2008}).

\bibitem{binnington-poisson}
\bibinfo{author}{Binnington, T.} \& \bibinfo{author}{Poisson, E.}
\newblock \bibinfo{journal}{\bibinfo{title}{{Relativistic theory of tidal Love
  numbers}}}.
\newblock {\emph{\JournalTitle{Phys.Rev.}}} \textbf{\bibinfo{volume}{D80}},
  \bibinfo{pages}{084018}, \doiprefix\url{10.1103/PhysRevD.80.084018}
  (\bibinfo{year}{2009}).
\newblock \eprint{0906.1366}.

\bibitem{damour-nagar}
\bibinfo{author}{Damour, T.} \& \bibinfo{author}{Nagar, A.}
\newblock \bibinfo{journal}{\bibinfo{title}{{Relativistic tidal properties of
  neutron stars}}}.
\newblock {\emph{\JournalTitle{Phys.Rev.}}} \textbf{\bibinfo{volume}{D80}},
  \bibinfo{pages}{084035}, \doiprefix\url{10.1103/PhysRevD.80.084035}
  (\bibinfo{year}{2009}).
\newblock \eprint{0906.0096}.

\bibitem{Racine:2004hs}
\bibinfo{author}{Racine, E.} \& \bibinfo{author}{Flanagan, E.~E.}
\newblock \bibinfo{journal}{\bibinfo{title}{{Post-1-Newtonian equations of
  motion for systems of arbitrarily structured bodies}}}.
\newblock {\emph{\JournalTitle{Phys. Rev. D}}} \textbf{\bibinfo{volume}{71}},
  \bibinfo{pages}{044010}, \doiprefix\url{10.1103/PhysRevD.71.044010}
  (\bibinfo{year}{2005}).
\newblock \bibinfo{note}{[Erratum: Phys.Rev.D 88, 089903 (2013)]},
  \eprint{gr-qc/0404101}.

\bibitem{Carson:2019rjx}
\bibinfo{author}{Carson, Z.}, \bibinfo{author}{Chatziioannou, K.},
  \bibinfo{author}{Haster, C.-J.}, \bibinfo{author}{Yagi, K.} \&
  \bibinfo{author}{Yunes, N.}
\newblock \bibinfo{journal}{\bibinfo{title}{{Equation-of-state insensitive
  relations after GW170817}}}.
\newblock {\emph{\JournalTitle{Phys. Rev. D}}} \textbf{\bibinfo{volume}{99}},
  \bibinfo{pages}{083016}, \doiprefix\url{10.1103/PhysRevD.99.083016}
  (\bibinfo{year}{2019}).
\newblock \eprint{1903.03909}.

\bibitem{Kastaun:2019bxo}
\bibinfo{author}{Kastaun, W.} \& \bibinfo{author}{Ohme, F.}
\newblock \bibinfo{journal}{\bibinfo{title}{{Finite tidal effects in GW170817:
  Observational evidence or model assumptions?}}}
\newblock {\emph{\JournalTitle{Phys. Rev. D}}} \textbf{\bibinfo{volume}{100}},
  \bibinfo{pages}{103023}, \doiprefix\url{10.1103/PhysRevD.100.103023}
  (\bibinfo{year}{2019}).
\newblock \eprint{1909.12718}.

\bibitem{PhysRev.116.1027}
\bibinfo{author}{Buchdahl, H.~A.}
\newblock \bibinfo{journal}{\bibinfo{title}{General relativistic fluid
  spheres}}.
\newblock {\emph{\JournalTitle{Phys. Rev.}}} \textbf{\bibinfo{volume}{116}},
  \bibinfo{pages}{1027--1034}, \doiprefix\url{10.1103/PhysRev.116.1027}
  (\bibinfo{year}{1959}).

\bibitem{Monitor:2017mdv}
\bibinfo{author}{Abbott, B.~P.} \emph{et~al.}
\newblock \bibinfo{journal}{\bibinfo{title}{{Gravitational Waves and Gamma-rays
  from a Binary Neutron Star Merger: GW170817 and GRB 170817A}}}.
\newblock {\emph{\JournalTitle{Astrophys. J.}}} \textbf{\bibinfo{volume}{848}},
  \bibinfo{pages}{L13}, \doiprefix\url{10.3847/2041-8213/aa920c}
  (\bibinfo{year}{2017}).
\newblock \eprint{1710.05834}.

\bibitem{Metzger:2019zeh}
\bibinfo{author}{Metzger, B.~D.}
\newblock \bibinfo{journal}{\bibinfo{title}{{Kilonovae}}}.
\newblock {\emph{\JournalTitle{Living Rev. Rel.}}}
  \textbf{\bibinfo{volume}{23}}, \bibinfo{pages}{1},
  \doiprefix\url{10.1007/s41114-019-0024-0} (\bibinfo{year}{2020}).
\newblock \eprint{1910.01617}.

\bibitem{2013PhRvL.111m1101B}
\bibinfo{author}{{Bauswein}, A.}, \bibinfo{author}{{Baumgarte}, T.~W.} \&
  \bibinfo{author}{{Janka}, H.~T.}
\newblock \bibinfo{journal}{\bibinfo{title}{{Prompt Merger Collapse and the
  Maximum Mass of Neutron Stars}}}.
\newblock {\emph{\JournalTitle{Phys. Rev. Lett.}}}
  \textbf{\bibinfo{volume}{111}}, \bibinfo{pages}{131101},
  \doiprefix\url{10.1103/PhysRevLett.111.131101} (\bibinfo{year}{2013}).
\newblock \eprint{1307.5191}.

\bibitem{2015ApJ...812...24F}
\bibinfo{author}{{Fryer}, C.~L.} \emph{et~al.}
\newblock \bibinfo{journal}{\bibinfo{title}{{The Fate of the Compact Remnant in
  Neutron Star Mergers}}}.
\newblock {\emph{\JournalTitle{Astrophys. J.}}} \textbf{\bibinfo{volume}{812}},
  \bibinfo{pages}{24}, \doiprefix\url{10.1088/0004-637X/812/1/24}
  (\bibinfo{year}{2015}).
\newblock \eprint{1504.07605}.

\bibitem{2015ApJ...808..186L}
\bibinfo{author}{{Lawrence}, S.}, \bibinfo{author}{{Tervala}, J.~G.},
  \bibinfo{author}{{Bedaque}, P.~F.} \& \bibinfo{author}{{Miller}, M.~C.}
\newblock \bibinfo{journal}{\bibinfo{title}{{An Upper Bound on Neutron Star
  Masses from Models of Short Gamma-Ray Bursts}}}.
\newblock {\emph{\JournalTitle{Astrophys. J.}}} \textbf{\bibinfo{volume}{808}},
  \bibinfo{pages}{186}, \doiprefix\url{10.1088/0004-637X/808/2/186}
  (\bibinfo{year}{2015}).
\newblock \eprint{1505.00231}.

\bibitem{2017ApJ...850L..19M}
\bibinfo{author}{{Margalit}, B.} \& \bibinfo{author}{{Metzger}, B.~D.}
\newblock \bibinfo{journal}{\bibinfo{title}{{Constraining the Maximum Mass of
  Neutron Stars from Multi-messenger Observations of GW170817}}}.
\newblock {\emph{\JournalTitle{Astrophys. J. Lett.}}}
  \textbf{\bibinfo{volume}{850}}, \bibinfo{pages}{L19},
  \doiprefix\url{10.3847/2041-8213/aa991c} (\bibinfo{year}{2017}).
\newblock \eprint{1710.05938}.

\bibitem{2017PhRvD..96l3012S}
\bibinfo{author}{{Shibata}, M.} \emph{et~al.}
\newblock \bibinfo{journal}{\bibinfo{title}{{Modeling GW170817 based on
  numerical relativity and its implications}}}.
\newblock {\emph{\JournalTitle{Phys. Rev. D}}} \textbf{\bibinfo{volume}{96}},
  \bibinfo{pages}{123012}, \doiprefix\url{10.1103/PhysRevD.96.123012}
  (\bibinfo{year}{2017}).
\newblock \eprint{1710.07579}.

\bibitem{2018ApJ...852L..25R}
\bibinfo{author}{{Rezzolla}, L.}, \bibinfo{author}{{Most}, E.~R.} \&
  \bibinfo{author}{{Weih}, L.~R.}
\newblock \bibinfo{journal}{\bibinfo{title}{{Using Gravitational-wave
  Observations and Quasi-universal Relations to Constrain the Maximum Mass of
  Neutron Stars}}}.
\newblock {\emph{\JournalTitle{Astrophys. J. Lett.}}}
  \textbf{\bibinfo{volume}{852}}, \bibinfo{pages}{L25},
  \doiprefix\url{10.3847/2041-8213/aaa401} (\bibinfo{year}{2018}).
\newblock \eprint{1711.00314}.

\bibitem{2018PhRvD..97b1501R}
\bibinfo{author}{{Ruiz}, M.}, \bibinfo{author}{{Shapiro}, S.~L.} \&
  \bibinfo{author}{{Tsokaros}, A.}
\newblock \bibinfo{journal}{\bibinfo{title}{{GW170817, general relativistic
  magnetohydrodynamic simulations, and the neutron star maximum mass}}}.
\newblock {\emph{\JournalTitle{Phys. Rev. D}}} \textbf{\bibinfo{volume}{97}},
  \bibinfo{pages}{021501}, \doiprefix\url{10.1103/PhysRevD.97.021501}
  (\bibinfo{year}{2018}).
\newblock \eprint{1711.00473}.

\bibitem{2019ApJ...887L..24M}
\bibinfo{author}{{Miller}, M.~C.} \emph{et~al.}
\newblock \bibinfo{journal}{\bibinfo{title}{{PSR J0030+0451 Mass and Radius
  from NICER Data and Implications for the Properties of Neutron Star
  Matter}}}.
\newblock {\emph{\JournalTitle{Astrophys. J. Lett.}}}
  \textbf{\bibinfo{volume}{887}}, \bibinfo{pages}{L24},
  \doiprefix\url{10.3847/2041-8213/ab50c5} (\bibinfo{year}{2019}).
\newblock \eprint{1912.05705}.

\bibitem{2021arXiv210506979M}
\bibinfo{author}{{Miller}, M.~C.} \emph{et~al.}
\newblock \bibinfo{journal}{\bibinfo{title}{{The Radius of PSR J0740+6620 from
  NICER and XMM-Newton Data}}}.
\newblock {\emph{\JournalTitle{arXiv e-prints}}}
  \bibinfo{pages}{arXiv:2105.06979} (\bibinfo{year}{2021}).
\newblock \eprint{2105.06979}.

\bibitem{2019ApJ...887L..21R}
\bibinfo{author}{{Riley}, T.~E.} \emph{et~al.}
\newblock \bibinfo{journal}{\bibinfo{title}{{A NICER View of PSR J0030+0451:
  Millisecond Pulsar Parameter Estimation}}}.
\newblock {\emph{\JournalTitle{Astrophys. J. Lett.}}}
  \textbf{\bibinfo{volume}{887}}, \bibinfo{pages}{L21},
  \doiprefix\url{10.3847/2041-8213/ab481c} (\bibinfo{year}{2019}).
\newblock \eprint{1912.05702}.

\bibitem{2021arXiv210506980R}
\bibinfo{author}{{Riley}, T.~E.} \emph{et~al.}
\newblock \bibinfo{journal}{\bibinfo{title}{{A NICER View of the Massive Pulsar
  PSR J0740+6620 Informed by Radio Timing and XMM-Newton Spectroscopy}}}.
\newblock {\emph{\JournalTitle{arXiv e-prints}}}
  \bibinfo{pages}{arXiv:2105.06980} (\bibinfo{year}{2021}).
\newblock \eprint{2105.06980}.

\bibitem{Bauswein:2020ggy}
\bibinfo{author}{Bauswein, A.} \& \bibinfo{author}{Blacker, S.}
\newblock \bibinfo{journal}{\bibinfo{title}{{Impact of quark deconfinement in
  neutron star mergers and hybrid star mergers}}}.
\newblock {\emph{\JournalTitle{Eur. Phys. J. ST}}}
  \textbf{\bibinfo{volume}{229}}, \bibinfo{pages}{3595--3604},
  \doiprefix\url{10.1140/epjst/e2020-000138-7} (\bibinfo{year}{2020}).
\newblock \eprint{2006.16183}.

\bibitem{Blaschke:2020vuy}
\bibinfo{author}{Blaschke, D.} \& \bibinfo{author}{Cierniak, M.}
\newblock \bibinfo{journal}{\bibinfo{title}{{Studying the onset of
  deconfinement with multi-messenger astronomy of neutron stars}}}.
\newblock {\emph{\JournalTitle{Astron. Nachr.}}}
  \textbf{\bibinfo{volume}{342}}, \bibinfo{pages}{227--233},
  \doiprefix\url{10.1002/asna.202113909} (\bibinfo{year}{2021}).
\newblock \eprint{2012.15785}.

\end{thebibliography}

\end{document}